\documentclass[epj,nopacs,fleqn]{svjour}
\pdfoutput=1
\usepackage{graphicx,amsmath,amssymb,slashed,cite}

\newcommand{\nn}{\nonumber}
\newcommand{\perc}{\%}
\newcommand\sss{\scriptscriptstyle}

\begin{document}

\title{Higgs characterisation via vector-boson fusion and associated
       production: NLO and parton-shower effects} 

\author{
 Fabio Maltoni\inst{1},
 Kentarou Mawatari\inst{2},
 Marco Zaro\inst{3}
}

\institute{ 
  Centre for Cosmology, Particle Physics and Phenomenology (CP3),
 Universit\'e Catholique de Louvain, \\
 B-1348 Louvain-la-Neuve, Belgium
 \and
 Theoretische Natuurkunde and IIHE/ELEM, Vrije Universiteit Brussel,
 and International Solvay Institutes,\\
 Pleinlaan 2, B-1050 Brussels, Belgium
 \and
 LPTHE, CNRS UMR 7589, UPMC Univ. Paris 6, Paris 75252, France
}

\abstract{
Vector-boson fusion and associated production at the LHC can provide key
information on the strength and structure of the Higgs couplings to the
Standard Model particles. Using an effective field theory approach, we
study the effects of next-to-leading order (NLO) QCD corrections matched
to parton shower on selected observables for various spin-0
hypotheses. We find that inclusion of NLO corrections is needed to
reduce the theoretical uncertainties on total rates as well as to
reliably predict the shapes of the distributions. Our results are
obtained in a fully automatic way via {\sc FeynRules} and
{\sc MadGraph5\_aMC@NLO}. 
}

\date{}

\titlerunning{Higgs characterisation via VBF and VH production}
\authorrunning{F.~Maltoni, K.~Mawatari, M.~Zaro}

\maketitle


\section{Introduction}\label{sec:intro}

After the discovery of a new boson at the
LHC~\cite{Aad:2012tfa,Chatrchyan:2012ufa}, studies of its properties
have become the first priority of the high energy physics community. A
coordinated theoretical and experimental effort is in
place~\cite{Dittmaier:2011ti,Dittmaier:2012vm,Heinemeyer:2013tqa} that
aims at maximising the information from the ongoing and forthcoming
measurements. On the experimental side, new analyses, strategies and
more precise measurements are being performed that cover the wider range
of relevant production and decay channels in the Standard Model (SM) and 
beyond, and the recent measurements of the coupling
strength~\cite{ATLAS:2013sla,Chatrchyan:2013lba} and the
spin-parity properties~\cite{Aad:2013xqa,Chatrchyan:2012jja} give strong
indication that the new particle is indeed the scalar boson predicted by
the SM. On the theoretical side, predictions for signal and background
are being obtained at higher orders in perturbative expansion in QCD and 
electroweak (EW), so that a better accuracy in the extraction of the SM
parameters can be achieved. In addition, new variables and observables
are being proposed that can be sensitive to new physics effects.  At the
same time, considerable attention is being devoted to the definition of
a theoretical methodology and framework to collect and interpret the
constraints coming from the experimental side.

The proposal of employing an effective field theory (EFT) that features
only SM particles and symmetries at the EW scale has turned out to be
particularly appealing. Such a minimal assumption, certainly well
justified by the present data, provides not only a drastic reduction of
all possible interactions that Lorentz symmetry alone would allow, but
also a well-defined and powerful framework where constraints coming from
Higgs measurements can be globally analysed together with those coming
from precision EW measurements and flavor physics (see for example
refs.~\cite{Hagiwara:1993qt,Giudice:2007fh,Gripaios:2009pe,Lafaye:2009vr,Low:2009di,Morrissey:2009tf,Contino:2010mh,Espinosa:2010vn,Azatov:2012bz,Espinosa:2012ir,Ellis:2012rx,Klute:2012pu,Low:2012rj,Corbett:2012dm,Ellis:2012hz,Montull:2012ik,Espinosa:2012im,Carmi:2012in,Plehn:2012iz,Passarino:2012cb,Corbett:2012ja,Cheung:2013kla,Falkowski:2013dza,Contino:2013kra,Chen:2013ejz},
and more in general refs.~\cite{Buchmuller:1985jz,Grzadkowski:2010es}). 

In this context, the {\it Higgs Characterisation} (HC)
framework has been recently presented~\cite{Artoisenet:2013puc} that
follows the general strategy outlined in ref.~\cite{Christensen:2009jx}.  
A simple EFT lagrangian featuring bosons with various
spin-parity assignments  has been implemented in 
{\sc FeynRules}~\cite{Christensen:2008py,Alloul:2013bka} and passed to the
{\sc MadGraph5\_aMC@NLO}~\cite{Alwall:2011uj,Frederix:2009yq,Hirschi:2011pa} 
framework by means of the {\sc UFO} model
file~\cite{Degrande:2011ua,deAquino:2011ub}. Such an implementation
is simple but general enough to describe any new physics effects coming
from higher scales in a fully model-independent way. It has the
advantage of being systematically and seamlessly improvable through the
inclusion of more operators in the lagrangian on one side and of
higher-order corrections, notably those coming from QCD, on the
other. The latter, considered in the form of multi-parton tree-level
computations (ME+PS) and of next-to-leading order (NLO) calculations
matched to parton showers (NLO+PS), are a very important ingredient for 
performing sensible phenomenological studies.  

In ref.~\cite{Artoisenet:2013puc} we have provided a study of higher
order QCD effects for inclusive $pp\to X(J^P)$ production, with 
$J^P=0^+$, $0^-$, $1^+$, $1^-,$ and $2^+$, and correlated decay of
resonances into a pair of gauge bosons, where gluon fusion ($q\bar q$
annihilation) is dominant for  
spin-0 and spin-2 (spin-1) at the LO. In this work, we present the
results for the next most important production channels at the LHC,
{\it i.e.}, weak vector-boson fusion (VBF) and associated production
(VH), focusing on the most likely spin-0 hypothesis. As already noted in 
ref.~\cite{Artoisenet:2013puc}, these processes share the 
property that NLO QCD corrections factorise exactly with respect to the
new physics interactions in Higgs couplings and therefore can be
automatically performed within the current {\sc MadGraph5\_aMC@NLO}
framework. Given that the Higgs characterisation can also be done
automatically in $t\bar tH$ production channel~\cite{Frederix:2011zi}, 
all the main Higgs production channels are covered. 

We stress that the spin-parity studies in VBF and VH production nicely
complement those in $H\to ZZ/WW$
decays~\cite{Hagiwara:2009wt,Englert:2012xt}.    
One of the advantages in the VBF and VH channels is that spin-parity
observables, {\it e.g.}, the azimuthal difference between the two
tagging jets $\Delta\phi_{jj}$ in VBF, do not require a reconstruction
of the Higgs resonance, although the separation between the $Z$ and $W$ 
contributions is very difficult. In this study, we focus on the effects
of the QCD corrections in Higgs VBF and VH production without
considering the decay.   

The paper is organised as follows.
In the following section we recall the relevant effective lagrangian of 
ref.~\cite{Artoisenet:2013puc}, and define the sample scenarios used to
illustrate the phenomenological implications.   
In sect.~\ref{sec:vbf} we present the VBF results in the form of
distributions of key observables in the inclusive setup as well as with 
dedicated VBF cuts, while in sect.~\ref{sec:vh} we illustrate the
$W^{\pm}H$ and $ZH$ production.  
We briefly summarise our findings in the concluding section.

\section{Theoretical setup}\label{sec:lag}

In this section, we summarise the full setup, from the lagrangian, to
the choice of benchmark scenarios, to event generation at NLO accuracy.

\subsection{Effective lagrangian and benchmark scenarios}

We construct an effective lagrangian below the electroweak symmetry
breaking (EWSB) scale in terms of mass eigenstates. Our assumptions are
simply that the resonance structure observed in data corresponds to one
bosonic state ($X(J^P)$ with $J=0$, $1$, or $2$, and a mass of about
$125$~GeV), and that no other new state below the cutoff $\Lambda$
coupled to such a resonance exists. We also follow the principle that
any new physics is dominantly described by the lowest dimensional
operators. This means, for the spin-0 case, that we include all effects
coming from the complete set of dimension-six operators with respect to
the SM gauge symmetry.

\begin{table}[b]
\center
\begin{tabular}{lll}
 \hline
 parameter\hspace*{5mm} & description \\
 \hline
 $\Lambda$ [GeV] & cutoff scale \\
 $c_{\alpha}\,(\equiv \cos\alpha$) & mixing between $0^+$ and
         $0^-$ \\
 $\kappa_i$ & dimensionless coupling parameter \\
 \hline
\end{tabular}
\caption{HC model parameters.}
\label{tab:param}
\end{table}

\begin{table}
\center
\begin{tabular}{cccccc}
 \hline
 $g_{Xyy'}\times v\ \ $ & $ZZ/WW$ &$\gamma\gamma$ & $Z\gamma$ \\
 \hline
 $X=H$ & $2m_{Z/W}^2$ & $47\alpha_{\sss\rm EM}/18\pi$ & $C(94c^2_W-13)/9\pi$ \\
 $X=A$ & 0        & $4\alpha_{\sss\rm EM}/3\pi$ & $2C(8c^2_W-5)/3\pi$ \\ 
 \hline
\end{tabular}
\caption{Values in units of $v$ taken by the couplings $g_{Xyy'}$ for
 the EW gauge bosons. $C=\sqrt{\frac{\alpha_{\sss\rm EM}G_F
 m_Z^2}{8\sqrt{2}\pi}}$.} 
\label{tab:gXaa}
\end{table}

The effective lagrangian relevant for this work, {\it i.e.}, for the
interactions between a spin-0 state and vector bosons, is (eq.~(2.4) in
ref.~\cite{Artoisenet:2013puc}): 
\begin{align}
 &{\cal L}_0^{V} =\bigg\{
  c_{\alpha}\kappa_{\rm SM}\big[\frac{1}{2}g_{\sss HZZ}\, Z_\mu Z^\mu 
                                +g_{\sss HWW}\, W^+_\mu W^{-\mu}\big] \nn\\
  &\quad -\frac{1}{4}\big[c_{\alpha}\kappa_{\sss H\gamma\gamma}
  g_{\sss H\gamma\gamma} \, A_{\mu\nu}A^{\mu\nu}
        +s_{\alpha}\kappa_{\sss A\gamma\gamma}g_{ \sss A\gamma\gamma}\,
  A_{\mu\nu}\widetilde A^{\mu\nu}
  \big] \nn\\
  &\quad -\frac{1}{2}\big[c_{\alpha}\kappa_{\sss HZ\gamma}g_{\sss HZ\gamma} \, 
  Z_{\mu\nu}A^{\mu\nu}
        +s_{\alpha}\kappa_{\sss AZ\gamma}g_{\sss AZ\gamma}\,Z_{\mu\nu}\widetilde A^{\mu\nu} \big] \nn\\
  &\quad -\frac{1}{4}\big[c_{\alpha}\kappa_{\sss Hgg}g_{\sss Hgg} \, G_{\mu\nu}^aG^{a,\mu\nu}
        +s_{\alpha}\kappa_{\sss Agg}g_{\sss Agg}\,G_{\mu\nu}^a\widetilde G^{a,\mu\nu} \big] \nn\\
  &\quad -\frac{1}{4}\frac{1}{\Lambda}\big[c_{\alpha}\kappa_{\sss HZZ} \, Z_{\mu\nu}Z^{\mu\nu}
        +s_{\alpha}\kappa_{\sss AZZ}\,Z_{\mu\nu}\widetilde Z^{\mu\nu} \big] \nn\\
  &\quad -\frac{1}{2}\frac{1}{\Lambda}\big[c_{\alpha}\kappa_{\sss HWW} \, W^+_{\mu\nu}W^{-\mu\nu}
        +s_{\alpha}\kappa_{\sss AWW}\,W^+_{\mu\nu}\widetilde W^{-\mu\nu}\big] \nn\\ 
  &\quad -\frac{1}{\Lambda}c_{\alpha} 
    \big[ \kappa_{\sss H\partial\gamma} \, A_{\nu}\partial_{\mu}A^{\mu\nu}
         +\kappa_{\sss H\partial Z} \, Z_{\nu}\partial_{\mu}Z^{\mu\nu} \nn\\
  &\hspace*{1.5cm}     + \, \big( \kappa_{\sss H\partial W} W_{\nu}^+\partial_{\mu}W^{-\mu\nu}+h.c.\big)
 \big]
 \bigg\} X_0  \,,
 \label{eq:0vv}
\end{align}
where the (reduced) field strength tensors are defined as
\begin{align}
 V_{\mu\nu} &=\partial_{\mu}V_{\nu}-\partial_{\nu}V_{\mu}\quad 
  (V=A,Z,W^{\pm})\,, \\
 G_{\mu\nu}^a &=\partial_{\mu}^{}G_{\nu}^a-\partial_{\nu}^{}G_{\mu}^a
  +g_sf^{abc}G_{\mu}^bG_{\nu}^c\,,
\end{align}
and the dual tensor is
\begin{align}
 \widetilde V_{\mu\nu}
 =\frac{1}{2}\epsilon_{\mu\nu\rho\sigma}V^{\rho\sigma}\,.
\end{align}
Our parametrisation: 
i) allows to recover the SM case easily by the dimensionless coupling
parameters $\kappa_i$ and the dimensionful couplings $g_{\sss Xyy'}$
shown in tables~\ref{tab:param} and \ref{tab:gXaa};  
ii) includes $0^-$ state couplings typical of SUSY or of generic
two-Higgs-doublet models (2HDM);
iii) describes $CP$-mixing between $0^+$ and $0^-$ states, parametrised
by an angle $\alpha$, in practice $-1<c_{\alpha}\,(\equiv\cos\alpha)<1$. 

The corresponding implementation of the dimension-six lagrangian above
the EWSB scale, where $SU(2)_L \times U(1)_Y$ is an exact symmetry, has
recently appeared~\cite{Alloul:2013naa} that has overlapping as well as
complementary features with respect to our HC lagrangian. We note that
the lagrangian of eq.~\eqref{eq:0vv} features 14 free parameters, of
which one possibly complex ($\kappa_{\sss H\partial W}$). On the other
hand, as explicitly shown in table~1 of ref.~\cite{Alloul:2013naa} these 
correspond to 11 free parameters in the parametrisation above the EWSB
due to the custodial symmetry. We stress that results at NLO in QCD
accuracy shown here can be obtained for that lagrangian in exactly the
same way.  

\begin{table}
\center
\begin{tabular}{ll}
\hline
 scenario & HC parameter choice \\ 
\hline
 $0^+$(SM) & $\kappa_{\sss \rm SM}=1\ (c_{\alpha}=1)$\\
 $0^+$(HD) &  $\kappa_{\sss HZZ,HWW}=1\ (c_{\alpha}=1)$\\
 $0^+$(HDder) & $\kappa_{\sss H\partial Z,H\partial W}=1\ (c_{\alpha}=1)$\\
 $0^+$(SM+HD) & $\kappa_{\sss SM,HZZ,HWW}=1\ (c_{\alpha}=1,\, \Lambda=v)$\\
 $0^-$(HD) &  $\kappa_{\sss AZZ,AWW}=1\ (c_{\alpha}=0)$\\
 $0^{\pm}$(HD) & $\kappa_{\sss HZZ,AZZ,HWW,AWW}=1\ (c_{\alpha}=1/\sqrt{2})$\\
\hline 
\end{tabular}
\caption{Benchmark scenarios.}
\label{tab:scenarios}
\end{table}

In table~\ref{tab:scenarios} we list the representative scenarios that
we later use for illustration. 
The first corresponds to the SM.
The second scenario, $0^+$(HD), includes only the $CP$-even higher
dimensional operators corresponding to $\kappa_{\sss HZZ,HWW}$ in a
custodial invariant way for VBF. 
The third scenario, $0^+$ (HDder), includes the so-called derivative
operators which, via the equations of motions, can be linked to contact
operators of the type $HVff'$.
The fourth scenario, $0^+$(SM+HD), features the interference, which
scales as $1/\Lambda$ in the physical observables, between the SM and
the HD operators.  
The fifth scenario, $0^-$(HD), is the analogous of the second one, but
for a pseudoscalar.  
Finally, the sixth scenario, $0^\pm$(HD), is representative of a
$CP$-mixed case, where the scalar is a scalar/pseudoscalar state in
equal proportion.

\subsection{NLO corrections including parton-shower effects}

The {\sc MadGraph5\_aMC@NLO} framework is designed to automatically
perform the computation of tree-level and NLO cross sections, possibly
including their matching to parton showers and the merging of samples
with different parton multiplicities. Currently, the full automation is
available in a unique and self-contained framework based on 
{\sc MadGraph5}~\cite{Alwall:2011uj} for SM processes with NLO QCD
corrections. User intervention is limited to the input of physics
quantities, and after event generation, to the choice of observables to
be analysed. In ref.~\cite{Artoisenet:2013puc} results for gluon fusion
have been presented and compared to predictions coming from ME+PS
(MLM-$k_T$ merging~\cite{Mangano:2001xp,Alwall:2007fs,Alwall:2008qv})
and NLO +PS. Distributions were found compatible between the two
predictions. In this work we limit ourselves to NLO+PS results as
typical observables are inclusive in terms of extra radiation and such
calculations do also provide a reliable normalisation. 

{\sc aMC@NLO} implements matching of any NLO QCD computation 
with parton showers following the MC@NLO approach~\cite{Frixione:2002ik}. 
Two independent and modular parts are devoted to the computation 
of specific contributions to an NLO-matched computation: 
{\sc MadFKS}~\cite{Frederix:2009yq} takes care of the Born, the
real-emission amplitudes, and it also performs the
subtraction of the infrared singularities and the generation of 
the Monte Carlo subtraction terms, according to 
the FKS prescription~\cite{Frixione:1995ms,Frixione:1997np};  
{\sc MadLoop}~\cite{Hirschi:2011pa} computes the one-loop amplitudes, 
using the CutTools~\cite{Ossola:2007ax} implementation of the OPP
integrand-reduction method~\cite{Ossola:2006us}. The {\sc OpenLoops}
method~\cite{Cascioli:2011va} is also used for better performance. 
Once the process
of interest is specified by the user, the generation of the code is
fully automated. Basic information, however, must be available about
the model and the interactions of its particles with QCD partons. For
{\sc MadFKS} this amounts to the ordinary Feynman rules. For
{\sc MadLoop}, on the other hand, the Feynman rules, UV counterterms,
and special tree-level rules, so-called $R_2$, necessary to (and defined
by) the OPP method, should be provided. While Feynman rules are
automatically computed from a given lagrangian (via 
{\sc FeynRules}~\cite{Christensen:2008py,Alloul:2013bka}), this is not
yet possible for UV counterterms and $R_2$ rules. At this moment this
limitation hampers the automatic computation of NLO QCD corrections for
arbitrary processes in generic BSM models, including the HC model. The
processes considered in this paper, VBF and VH, are, however, a notable
exception as QCD corrections can be computed automatically and in full
generality. This is because the corresponding one-loop amplitudes only
include SM particles and do not need any UV counterterms and $R_2$
information from the HC lagrangian. In the case of VBF, this assumes
that only vertex loop-corrections can be computed, {\it i.e.}, the
pentagon diagrams are discarded as the contributions only affect
interferences between the diagrams, which are negligible already at LO.

\subsection{Simulation parameters}

In our simulations we generate events at the LHC with a
center-of-mass energy $\sqrt{s}=8$~TeV and set the resonance mass to
$m_{X_0}=125$~GeV. Parton distributions functions (PDFs) are evaluated by
using the MSTW2008 (LO/NLO) parametrisation~\cite{Martin:2009iq}, and
jets are reconstructed via the anti-$k_T$ ($\Delta R=0.4$)
algorithm~\cite{Cacciari:2008gp} as implemented in 
{\sc FastJet}~\cite{Cacciari:2011ma}. Central values for the
renormalisation and factorisation scales $\mu_{R,F} $ are set to
$\mu_0=m_W$ and $m_{VH}$ for VBF and VH production, respectively, where
$m_{VH}$ is the invariant mass of the VH system. We note here that
scale (and PDF) uncertainties can be evaluated automatically in the code
via a reweighting technique~\cite{Frederix:2011ss}, the user only
deciding the range of variation. In addition, such information is
available on an event-by-event basis and therefore uncertainty bands can
be plotted for any observable of interest. In this work, however, to
simplify the presentation that focuses on the differences between the
various scenarios, we give this information only for total cross
sections and  refrain from showing them in the differential
distributions. For parton shower and hadronisation we employ 
{\sc HERWIG6}~\cite{Corcella:2000bw} in this paper, while 
{\sc HERWIG++}~\cite{Bahr:2008pv}, (virtuality ordered)
{\sc Pythia6}~\cite{Sjostrand:2006za} and
{\sc Pythia8}~\cite{Sjostrand:2007gs} are available to use in
{\sc aMC@NLO}, and the comparison among the above different shower
schemes was done for the SM Higgs boson in VBF in
ref.~\cite{Frixione:2013mta}.

\section{Vector boson fusion}\label{sec:vbf}

Predictions for Higgs production via VBF in the SM are known up to NNLO
accuracy for the total cross
section~\cite{Harlander:2008xn,Bolzoni:2010xr,Bolzoni:2011cu}, at the NLO
QCD~\cite{Han:1992hr,Figy:2003nv,Berger:2004pca,Figy:2004pt,Hankele:2006ma,Figy:2010ct}
+ EW~\cite{Ciccolini:2007jr,Ciccolini:2007ec}
level in a differential way and at NLO in QCD plus parton shower both in
the {\sc POWHEG BOX}~\cite{Nason:2009ai} and in 
{\sc aMC@NLO}~\cite{Frixione:2013mta}. NLO QCD predictions that include 
anomalous couplings between the Higgs and a pair of vector bosons are
available in {\sc VBFNLO}~\cite{Arnold:2008rz,Arnold:2011wj}. Our
implementation provides the first predictions for EFT interactions
including NLO corrections in QCD interfaced with a parton shower. Many
phenomenological studies on Higgs spin, parity and couplings are
available in the 
literature~\cite{Plehn:2001nj,Hagiwara:2009wt,Englert:2012ct,Andersen:2012kn,Englert:2012xt,Djouadi:2013yb,Englert:2013opa,Frank:2013gca,Anderson:2013afp},
which could now be upgraded to NLO+PS accuracy.

In our framework the code and events for VBF can be automatically
generated by issuing the following commands (note the \textsc{\$\$} sign
to forbid diagrams with $W^{\pm}$ or $Z$ bosons in the $s$-channel which
are included in VH production):  
\begin{verbatim}
 > import model HC_NLO_X0
 > generate p p > x0 j j $$ w+ w- z QCD=0 [QCD]
 > output
 > launch
\end{verbatim}
As a result all processes featuring a $VV^\prime\to X_0$ vertex, with
$V=W,Z,\gamma$ are generated, therefore including $\gamma\gamma\to X_0$
and $Z\gamma\to X_0$. We do not investigate their effects in our
illustrative studies below ({\it i.e.}, we set the corresponding
$\kappa_i$ to zero in the simulation), as we focus on SM-like VBF
observables. As mentioned above, since our interest is geared towards
QCD effects on production distributions, we do not include Higgs decays
in our studies either. We stress, however,  that  decays (as predicted
in the HC model) can be efficiently included at the partonic event level 
(before passing the event to a shower program) via 
{\sc MadSpin}~\cite{Artoisenet:2012st}. 

\begin{table}
\center
\begin{tabular}{lrlrlc}
 \hline
 scenario    & $\sigma_{\rm LO}$ & \hspace*{-3mm}(fb) 
             & $\sigma_{\rm NLO}$ & \hspace*{-3mm}(fb) & $K$ \\
 \hline
 $0^+$(SM)   & 1509(1) & \hspace*{-3mm}${}^{+4.7\perc}_{-4.4\perc}$  
             & 1633(2) & \hspace*{-3mm}${}^{+2.0\perc }_{-1.5\perc}$ 
             & 1.08 \\
 $0^+$(HD)   & 69.66(6) & \hspace*{-3mm}${}^{+7.5\perc}_{-6.6\perc}$
             & 67.08(13)  & \hspace*{-3mm}${}^{+2.2\perc}_{-2.3\perc}$ 
             & 0.96 \\
 $0^+$(HDder)& 721.9(6) & \hspace*{-3mm}${}^{+11.0\perc}_{-9.0\perc}$  
             & 684.9(1.5) & \hspace*{-3mm}${}^{+2.3\perc}_{-2.8\perc}$ 
             & 0.95 \\
 $0^+$(SM+HD)& 3065(2) & \hspace*{-3mm}${}^{+5.6\perc}_{-5.1\perc}$  
             & 3144(5) & \hspace*{-3mm}${}^{+1.6\perc}_{-1.1\perc}$ 
             & 1.03 \\
 $0^-$(HD)   & 57.10(4) & \hspace*{-3mm}${}^{+7.7\perc}_{-6.7\perc}$  
             & 55.24(11) & \hspace*{-3mm}${}^{+2.1\perc}_{-2.5\perc}$ 
             & 0.97 \\
 $0^\pm$(HD) & 63.46(5) & \hspace*{-3mm}${}^{+7.6\perc}_{-6.7\perc}$  
             & 61.07(13) & \hspace*{-3mm}${}^{+2.3\perc}_{-2.0\perc}$ 
             & 0.96 \\
 \hline
\end{tabular}
\caption{VBF total cross sections with scale uncertainties and
 corresponding $K$-factors at LHC 8TeV for various scenarios.} 
\label{tab:xsecVBF}
\end{table}

In table~\ref{tab:xsecVBF}, we first collect results for total cross
sections at LO and NLO accuracy together with scale uncertainties and
corresponding $K$-factors for the six scenarios defined in
table~\ref{tab:scenarios}. We do not impose any cuts here, and hence the
cross sections are identical with and without parton shower. The cross
sections for the HD hypotheses are calculated with the corresponding
$\kappa_i$ set to one and the cutoff scale $\Lambda=1$~TeV except for
the $0^+$(SM+HD) scenario, where we set $\Lambda=v=246$~GeV. We do this
to allow for visible effects of the interference between the SM and HD
terms. Equivalently, we could have kept $\Lambda=1$~TeV and chosen a
larger value for $\kappa_i$, as only the ratio $\kappa_i/\Lambda$ is
physical. The figures in parentheses give the numerical integration
uncertainties in the last digit(s). The other uncertainties correspond
to the envelope obtained by varying independently the renormalisation
and factorisation scales around the central value
$1/2<\mu_{R,F}/\mu_{\rm 0}<2$ with $\mu_0=m_W$. NLO QCD corrections
contribute constructively for the SM case, while destructively for the
HD cases, 
although the global $K$-factors are rather mild. The uncertainty in the
HD scenarios, especially for the derivative operator (HDder), are larger
than that in the SM case. Manifestly, the uncertainties are
significantly reduced going from LO to NLO. 

For the studies on the distributions, we require the presence of at
least two reconstructed jets with 
\begin{align}
 p_T^j>25~{\rm GeV}\,,\quad |\eta^j|<4.5\,.
\label{mincut}
\end{align}
In addition, we simulate a dedicated VBF selection by imposing an
invariant mass cut on the two leading jets, 
\begin{align}
 m(j_1,j_2)>500~{\rm GeV}\,.
\label{mjjcut} 
\end{align}
As well known such a cut has the scope to minimise the contributions
from gluon fusion and allow to extract VBF couplings. We note that we
do not put the rapidity separation cut, although this is the common VBF
cut, since $\Delta\eta(j_1,j_2)$ itself is a powerful observable to
determine the $HVV$ structure in VBF
production~\cite{Englert:2012xt,Djouadi:2013yb}.  

\begin{figure}
 \center 
 \includegraphics[width=0.48\textwidth,clip, trim = 100 220 50 160]{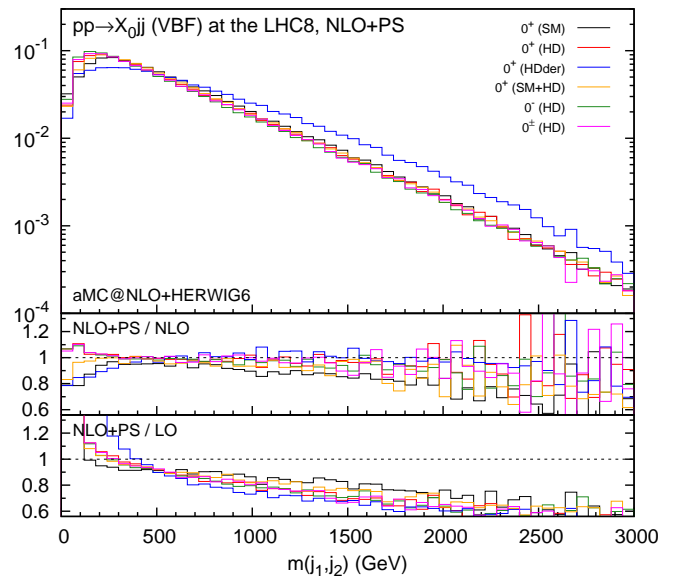}
 \caption{Distribution for the invariant mass of the two
 leading jets in VBF production with the acceptance cuts. The histograms
 in the main plot are normalized to unity.} 
\label{fig:mjj}
\end{figure} 

We start by showing the invariant mass distribution of the two leading 
jets in fig.~\ref{fig:mjj} for the six scenarios of
table~\ref{tab:scenarios}, where the minimal detector cuts in
eq.~\eqref{mincut} are applied. With the exception of the scenario
featuring the derivative operator (HDder), the distributions are all
very similar. This means that the invariant mass cut in
eq.~\eqref{mjjcut}, which is imposed in typical VBF selections, acts in
a similar way on all scenarios. 

The lowest inset in fig.~\ref{fig:mjj} is the ratio of NLO+PS to LO
results, while the middle one shows the ratio of NLO+PS to pure
NLO. NLO+PS corrections modify in consistent way LO parton-level
predictions with major effects at high invariant mass, {\it i.e.}, the
QCD corrections tend to make the tagging jets softer. In addition,
parton shower affects both the lower and higher invariant mass regions. 

\begin{figure*}
 \center 
 \includegraphics[width=0.48\textwidth,clip, trim = 90 210 50 160]{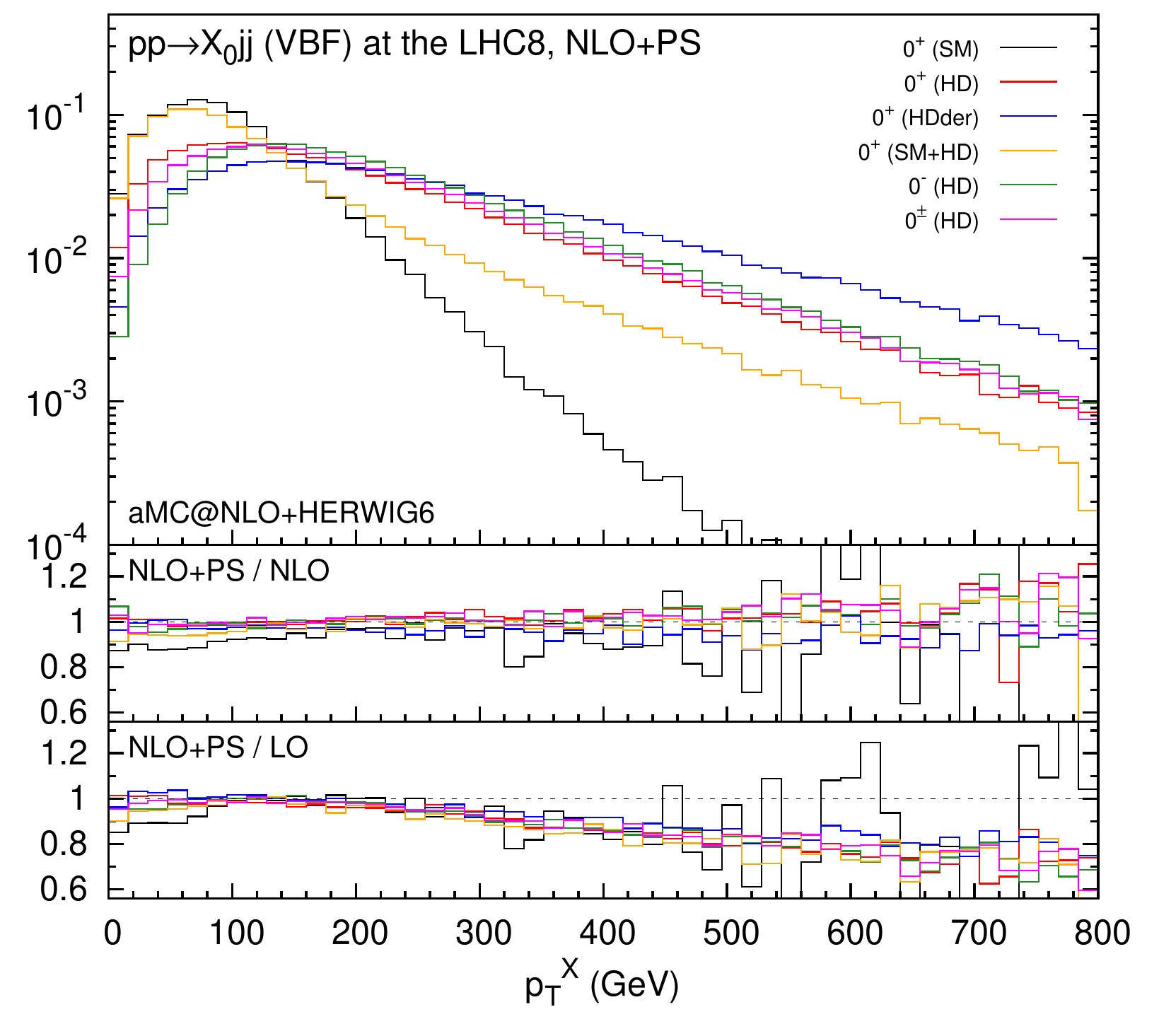}
 \includegraphics[width=0.48\textwidth,clip, trim = 90 210 50 160]{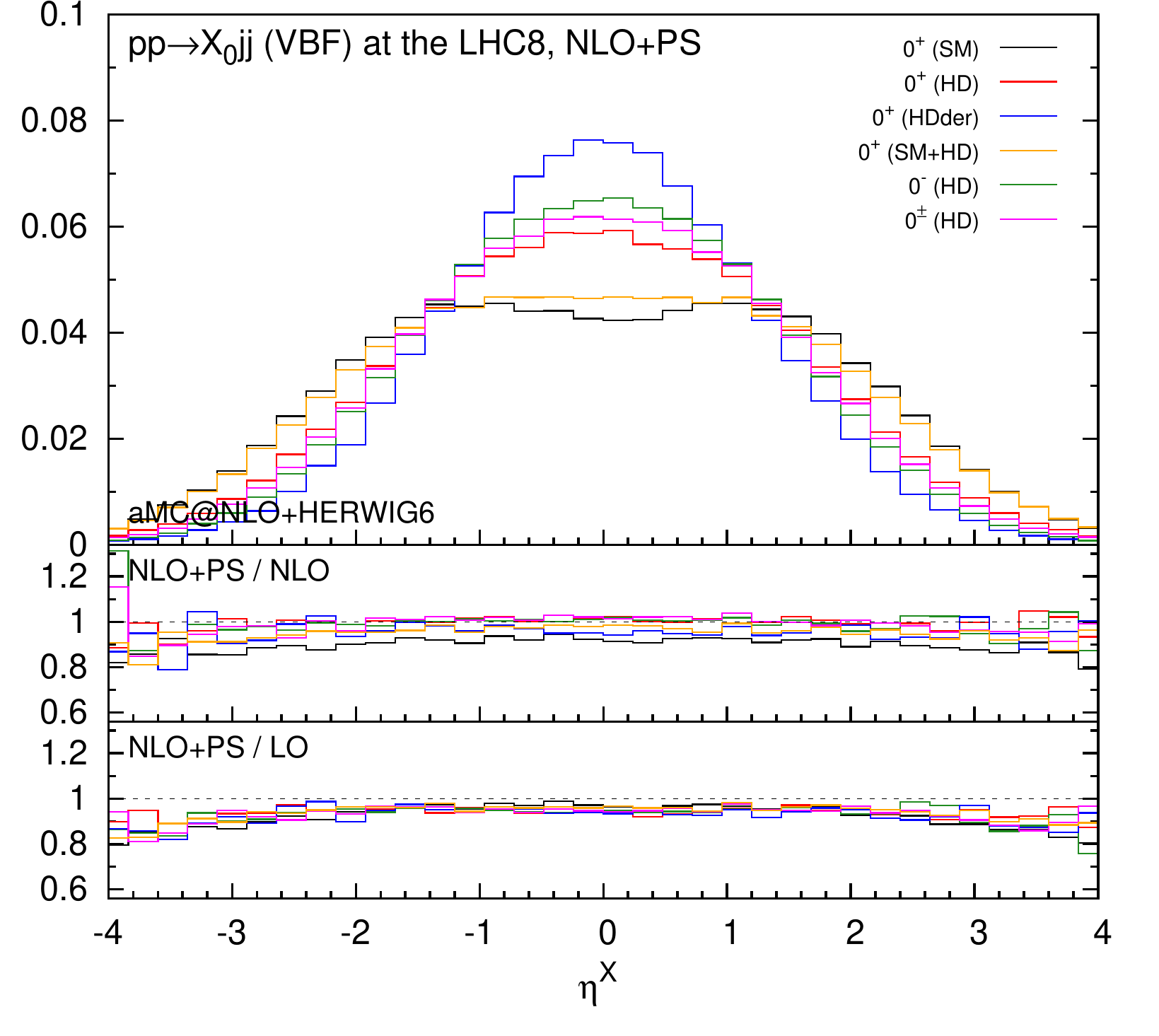}\\
 \includegraphics[width=0.48\textwidth,clip, trim = 90 210 50 160]{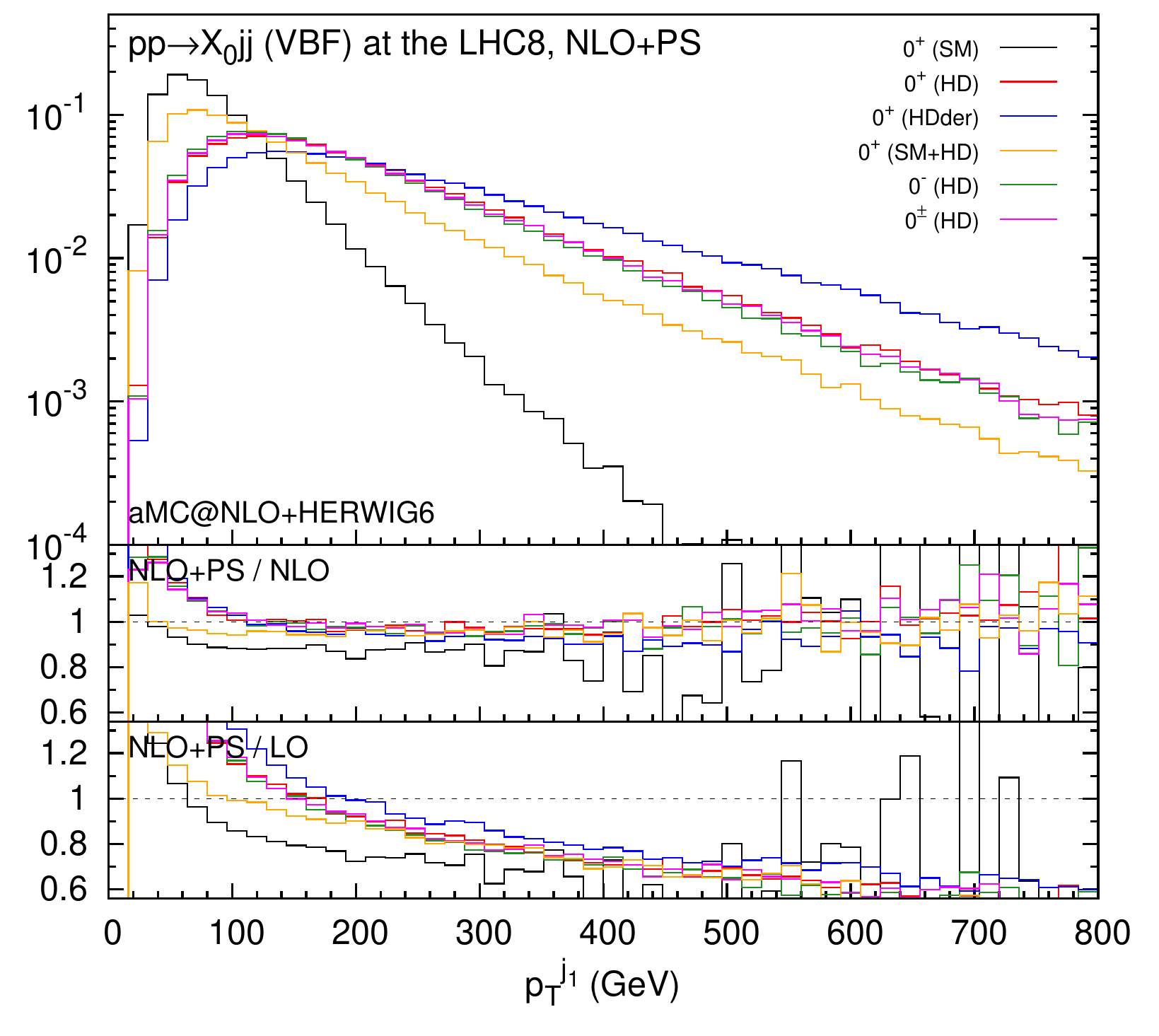}
 \includegraphics[width=0.48\textwidth,clip, trim = 90 210 50 160]{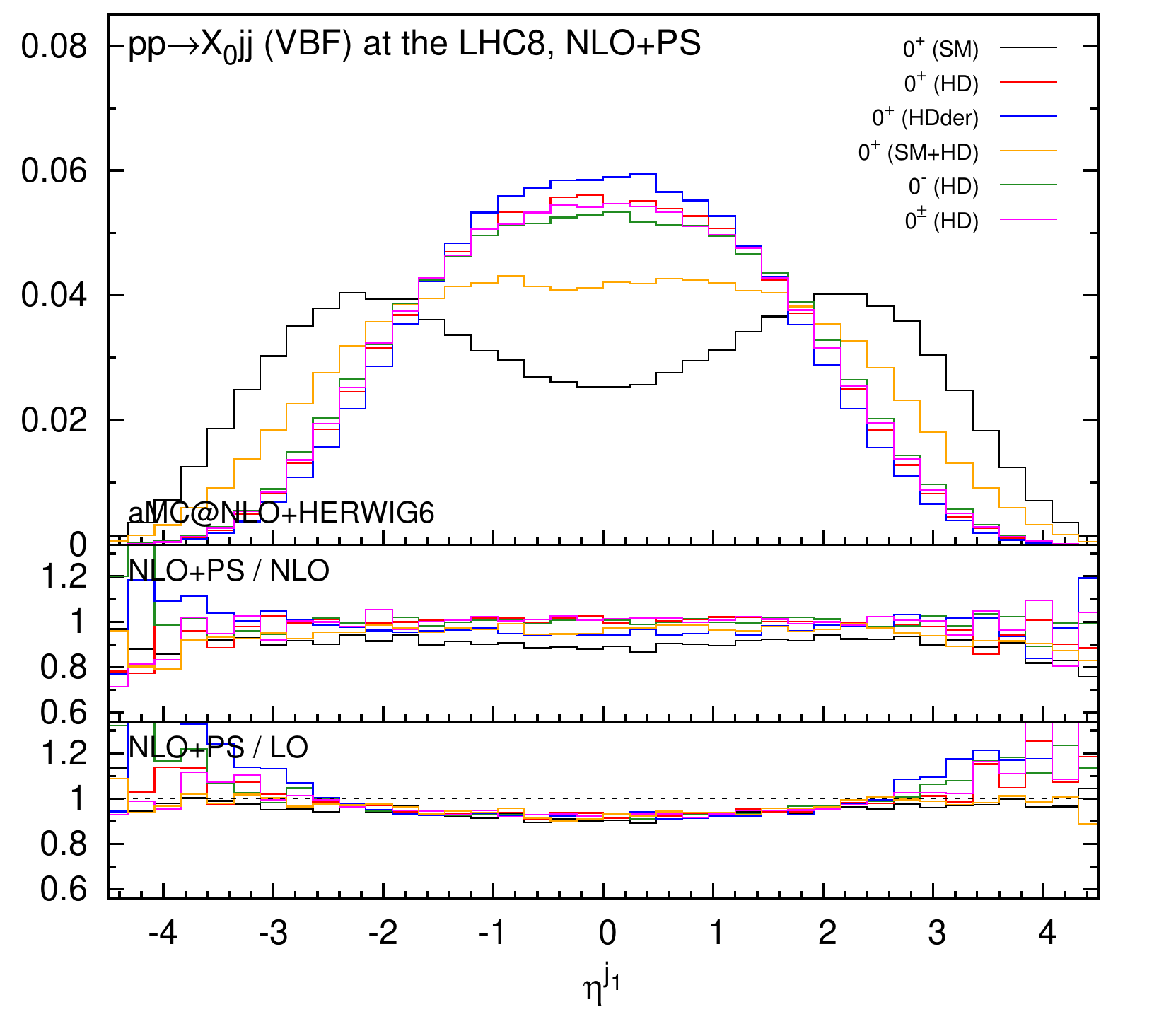}\\
 \includegraphics[width=0.48\textwidth,clip, trim = 90 210 50 160]{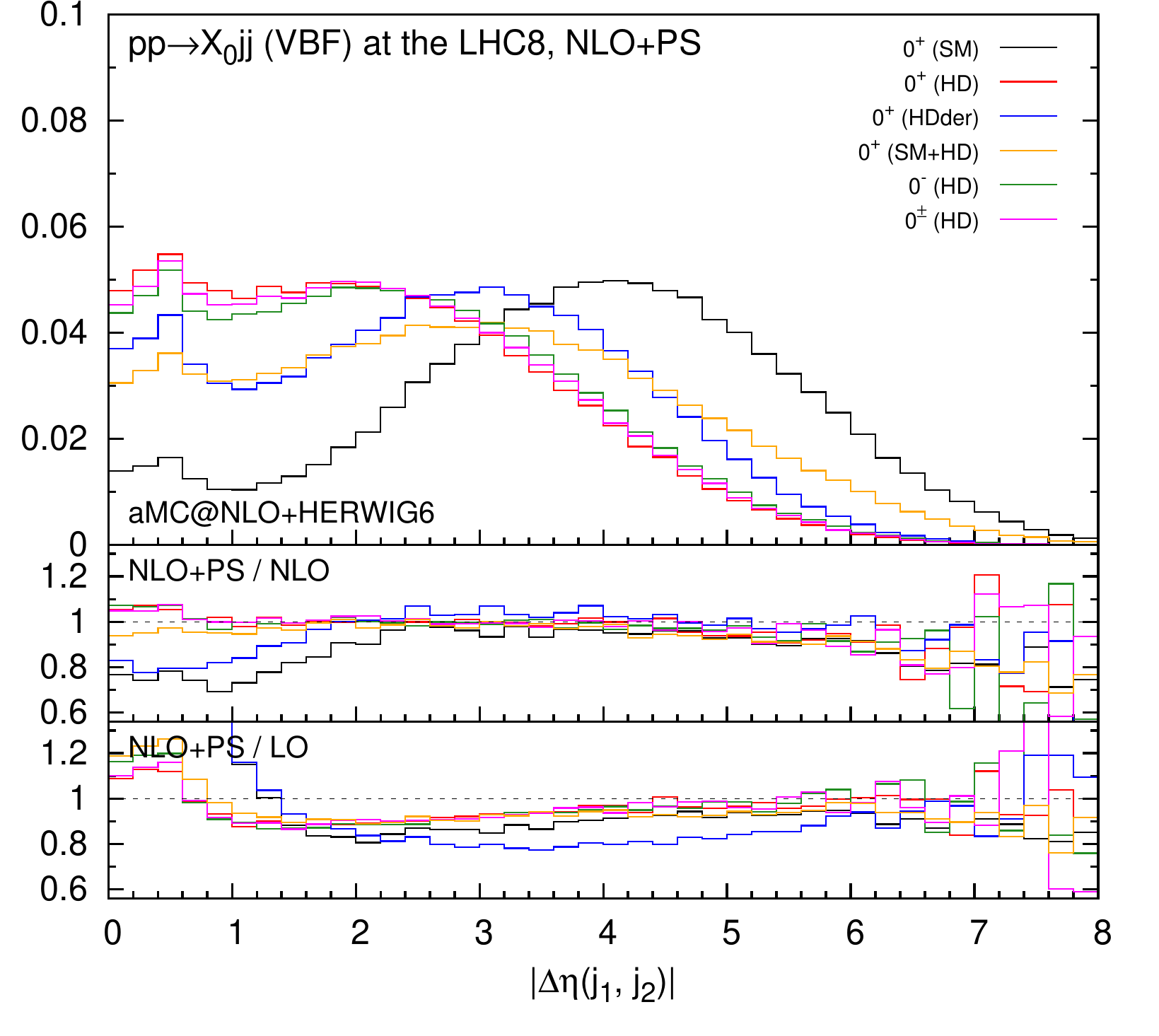}
 \includegraphics[width=0.48\textwidth,clip, trim = 90 210 50 160]{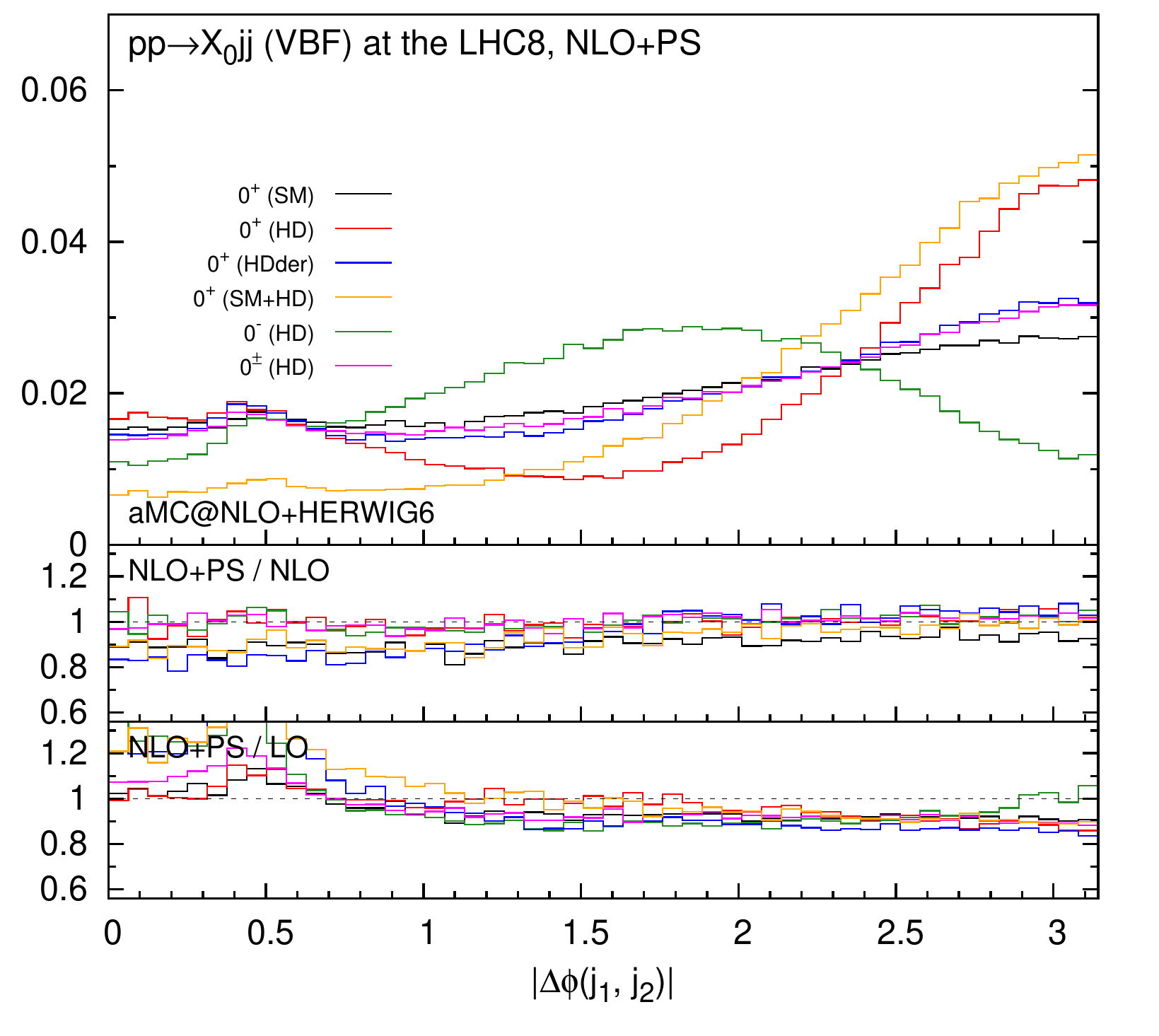}
 \caption{Distributions for $p_T^X$, $\eta^X$, $p_T^{j_1}$,
 $\eta^{j_1}$, $\Delta\eta(j_1,j_2)$, and $\Delta\phi(j_1,j_2)$ in VBF
 with the acceptance cuts for the jets. The histograms
 in the main plots are normalized to unity.} 
\label{fig:vbf1}
\end{figure*} 

\begin{figure*}
 \center 
 \includegraphics[width=0.48\textwidth,clip, trim = 90 210 50 160]{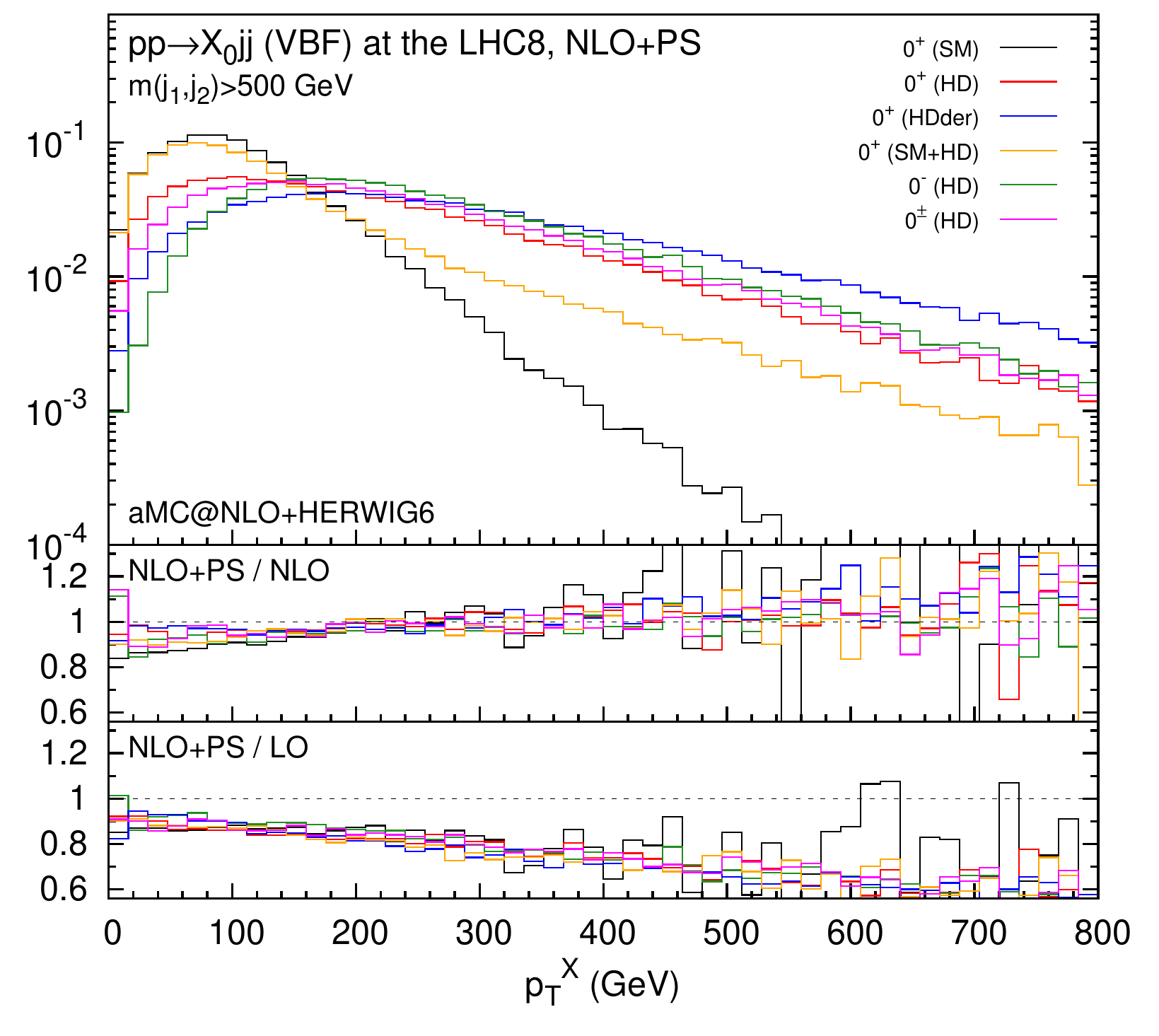}
 \includegraphics[width=0.48\textwidth,clip, trim = 90 210 50 160]{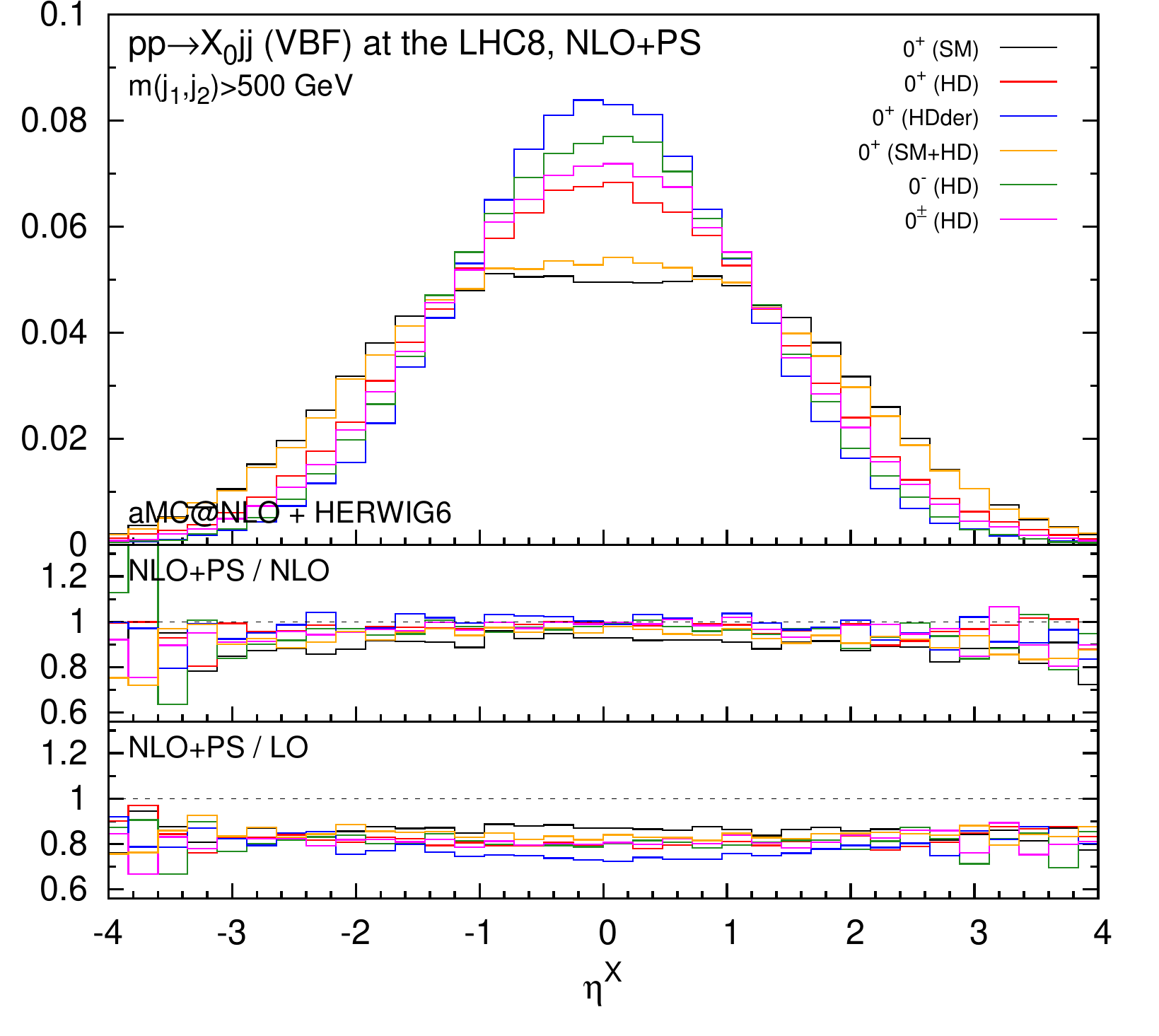}\\
 \includegraphics[width=0.48\textwidth,clip, trim = 90 210 50 160]{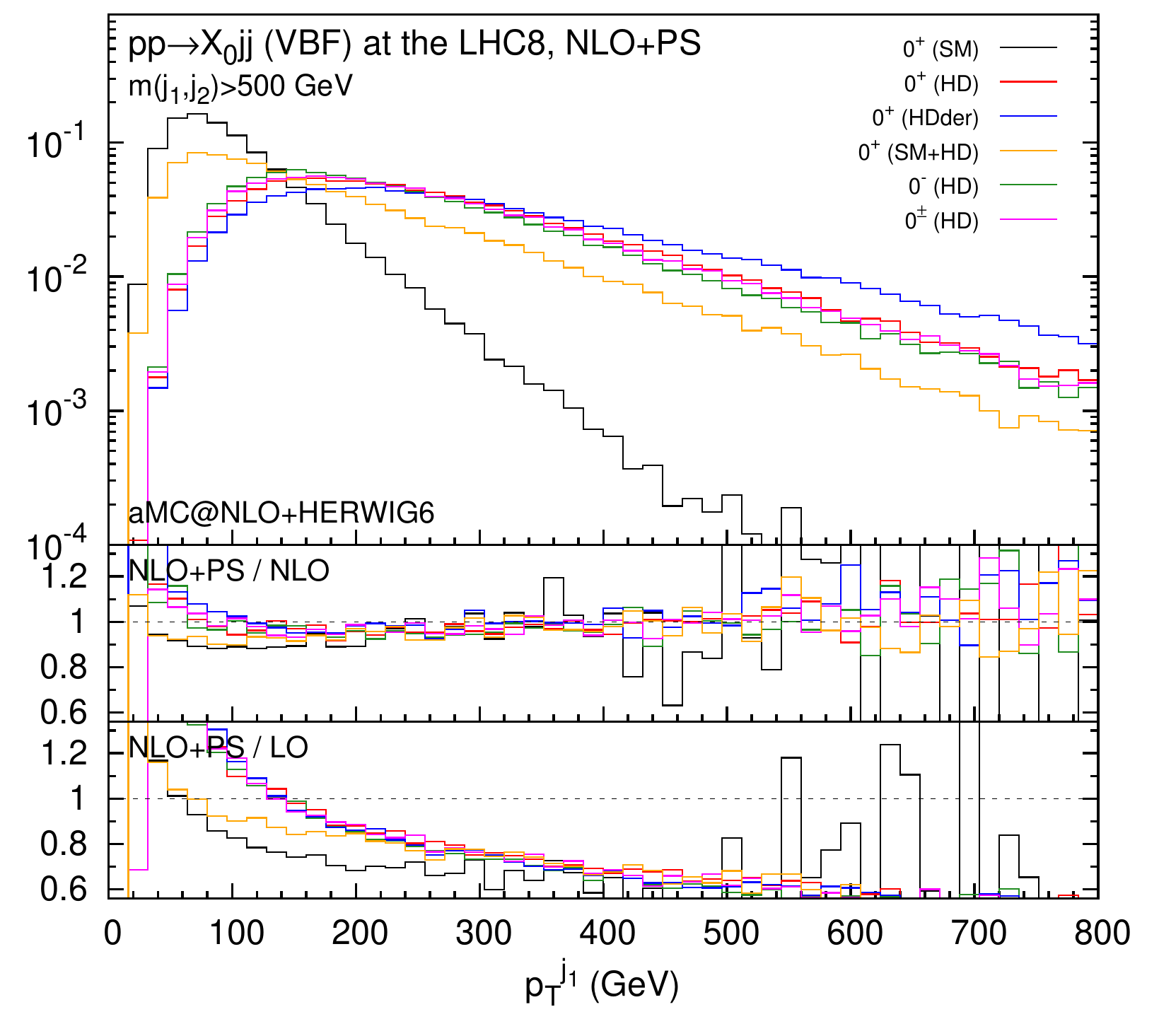}
 \includegraphics[width=0.48\textwidth,clip, trim = 90 210 50 160]{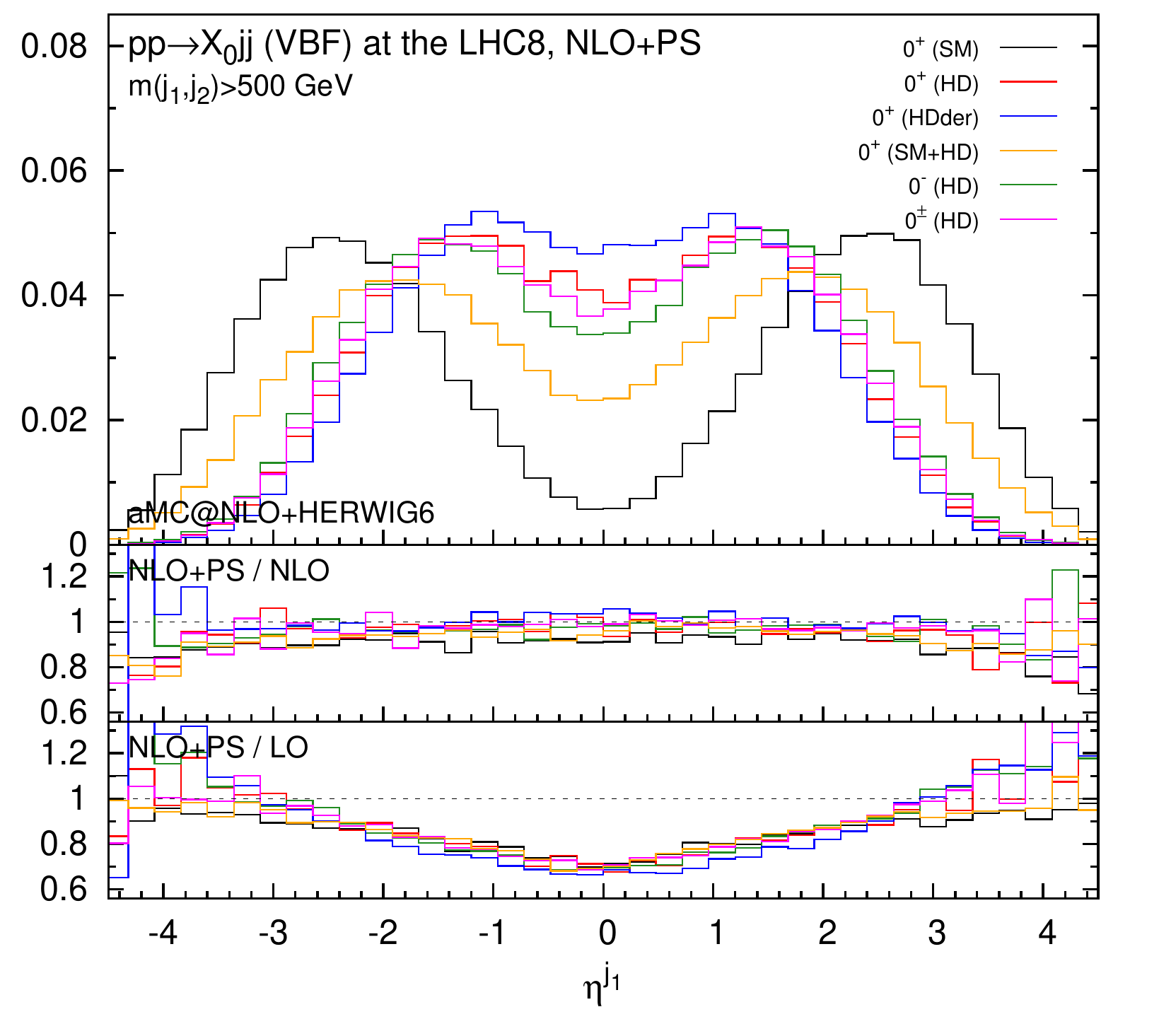}\\
 \includegraphics[width=0.48\textwidth,clip, trim = 90 210 50 160]{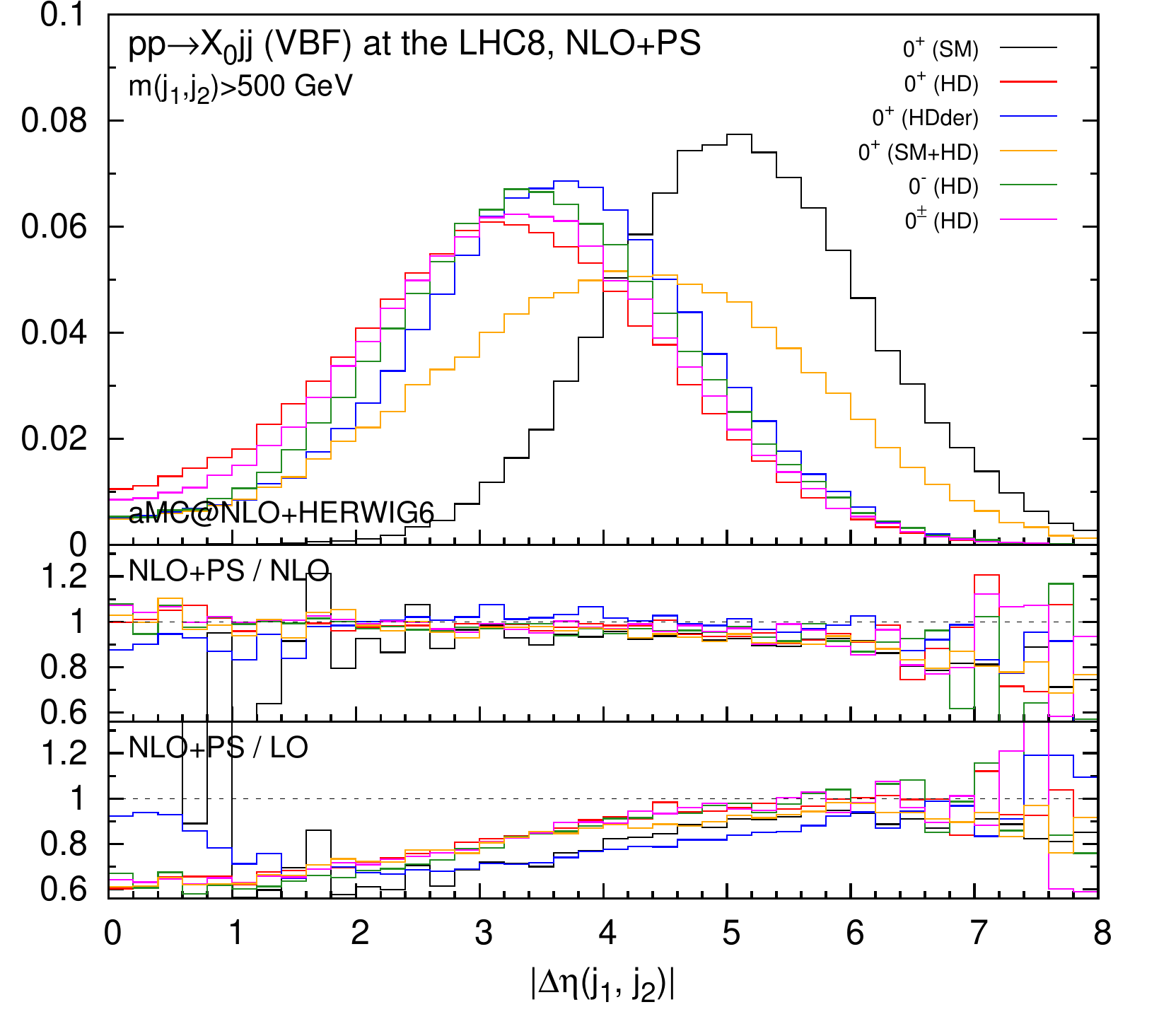}
 \includegraphics[width=0.48\textwidth,clip, trim = 90 210 50 160]{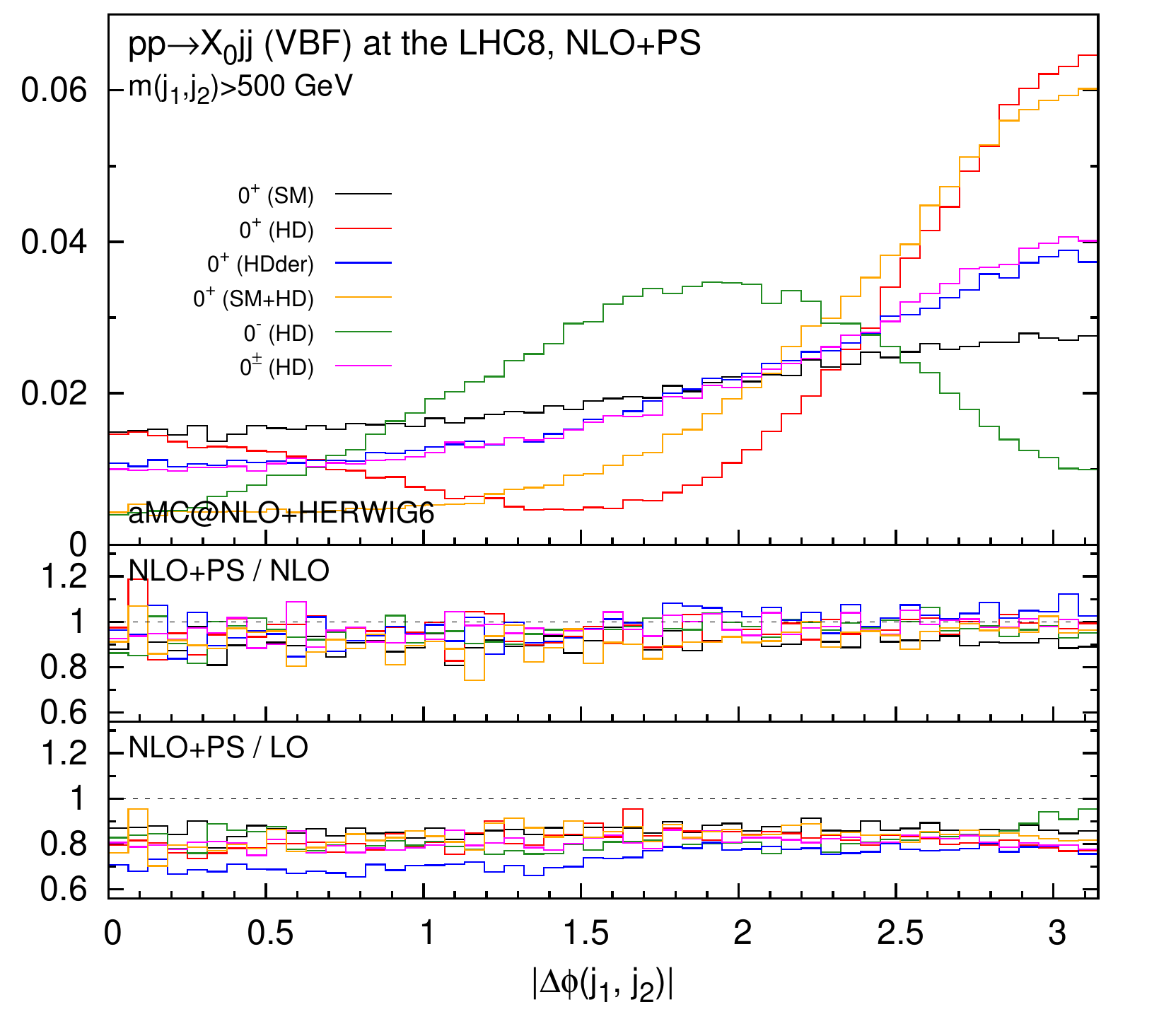} 
 \caption{Same as fig.~\ref{fig:vbf1}, but with the additional VBF cut in
 eq.~\eqref{mjjcut}.}  
\label{fig:vbf2}
\end{figure*} 

Figures~\ref{fig:vbf1} and \ref{fig:vbf2} collect key plots for the
$X_0$ and the hardest jet distributions, as well as the rapidity and
azimuthal separation of the two leading jets. In fig.~\ref{fig:vbf1}
only the acceptance cuts in eq.~\eqref{mincut} are imposed, while in
fig.~\ref{fig:vbf2} the additional VBF cut in eq.~\eqref{mjjcut} is
applied. As one can see, the invariant mass cut effectively suppresses
the central jet activity, especially for the SM case, while the
difference of the distributions among the different scenarios becomes 
more pronounced.
 
The unitarity violating behaviour of the higher dimensional
interactions, especially for $0^+$(HDder), clearly manifests itself in
the transverse momentum distributions for the $X_0$ and the jets.  The
rapidity distribution of the tagging jets displays the fact that in the
case of higher-dimensional interactions the jets result to be much more
central than the SM case. Same glaring difference appear in the
azimuthal correlations between the jets which offer clear handle to
discriminate about different interactions type and parity assignments. 

In all cases NLO corrections are relevant and cannot be described by an
overall $K$-factor. Moreover, their impact depends on the applied
cuts. Apart from regions in phase space where the jets end up close and
therefore are sensitive to NLO/jet reconstruction effects, the 
parton-shower effect on the shapes is very minor, especially after the
VBF cut.

\section{Vector boson associated production}\label{sec:vh}

Predictions for Higgs production in association with a weak vector boson
in the SM are known up to NNLO
accuracy~\cite{Brein:2003wg,Ferrera:2011bk,Brein:2012ne}, including EW
corrections~\cite{Ciccolini:2003jy,Denner:2011id}. NLO+PS results can be
obtained via {\sc (a)MC@NLO}~\cite{Frixione:2005gz,LatundeDada:2009rr}
and the {\sc POWHEG BOX}~\cite{Luisoni:2013cuh}. Many phenomenological
studies on Higgs spin, parity and couplings are available in the
literature~\cite{Miller:2001bi,Christensen:2010pf,Desai:2011yj,Ellis:2012xd,Englert:2012xt,Ellis:2013ywa,Godbole:2013saa,Isidori:2013cga,Delaunay:2013npa,Anderson:2013afp}. 
In this section we present the first predictions for EFT interactions
including NLO corrections in QCD interfaced with a parton shower in the
VH process. 

The code and events for VH production at hadron colliders can be
automatically generated by issuing the following commands:  
\begin{verbatim}
 > import model HC_NLO_X0
 > generate p p > x0 e+ ve [QCD]
 > add process p p > x0 e- ve~ [QCD]
 > add process p p > x0 e+ e- [QCD]
 > output
 > launch
\end{verbatim}
Note that the $W,Z$ decays are performed at the level of the matrix
elements and therefore all spin correlations are kept exactly. Again, as
in sect.~\ref{sec:vbf}, we do not consider contributions involving the
$X_0\gamma\gamma$ and $X_0Z\gamma$ vertices. 

Results for total cross sections (without any cuts) at LO and NLO
accuracy and corresponding $K$-factors for the six scenarios defined in
table~\ref{tab:scenarios} are collected in tables~\ref{tab:xsecWpH},
\ref{tab:xsecWmH} and \ref{tab:xsecZH} for $pp\to W^+H$, $W^-H$, and
$ZH$, respectively, including the $W/Z$ decay branching ratio into a
lepton pair. As in the VBF case, the uncertainties correspond to the
envelope of independently varying the renormalisation and factorisation
scales around the central value $1/2<\mu_{R,F}/\mu_{\rm 0}<2$ with
$\mu_0=m_{VH}$. Apart from the case of the SM for which the
uncertainties are accidentally small at LO, the results at NLO display
an improved stability. Quite interestingly all $K$-factors are found to
be around 1.3 for all the scenarios, with tiny difference among the
processes due to the different initial states. We note that the
cancellation of the $s$-channel vector-boson propagator due to the
derivative in the higher-dimensional scenarios results in the rather
large cross section in spite of the $\Lambda=1$~TeV cutoff (except for
the $0^+$(SM+HD) scenario, where $\Lambda=v=246$~GeV).   

\begin{table}
\center
\begin{tabular}{lrlrlc}
 \hline
 scenario    & $\sigma_{\rm LO}$ & \hspace*{-3mm}(fb) 
             & $\sigma_{\rm NLO}$ & \hspace*{-3mm}(fb) & $K$ \\
 \hline
 $0^+$(SM)   & 39.58(3) & \hspace*{-3mm}${}^{+0.1\perc}_{-0.6\perc}$  
             & 51.22(5) & \hspace*{-3mm}${}^{+2.2\perc}_{-1.8\perc }$ 
             & 1.29 \\
 $0^+$(HD)   & 13.51(1) & \hspace*{-3mm}${}^{+1.5\perc}_{-1.7\perc}$  
             & 17.51(1) & \hspace*{-3mm}${}^{+1.9\perc}_{-1.3\perc}$ 
             & 1.30 \\
 $0^+$(HDder)& 324.2(2) & \hspace*{-3mm}${}^{+4.7\perc}_{-4.3\perc}$  
             & 416.1(4) & \hspace*{-3mm}${}^{+2.3\perc}_{-2.1\perc}$ 
             & 1.28 \\
 $0^+$(SM+HD)& 118.8(1) & \hspace*{-3mm}${}^{+3.0\perc}_{-2.9\perc}$  
             & 154.2(1) & \hspace*{-3mm}${}^{+1.8\perc}_{-1.6\perc}$ 
             & 1.30 \\
 $0^-$(HD)   & 8.386(7) & \hspace*{-3mm}${}^{+2.6\perc}_{-2.6\perc}$  
             & 10.89(1) & \hspace*{-3mm}${}^{+1.8\perc}_{-1.5\perc}$ 
             & 1.30 \\
 $0^\pm$(HD) & 10.96(1) & \hspace*{-3mm}${}^{+1.9\perc}_{-2.1\perc}$  
             & 14.22(1) & \hspace*{-3mm}${}^{+1.8\perc}_{-1.3\perc}$ 
             & 1.30 \\
 \hline
\end{tabular}
\caption{$pp\to H(W^+\to e^+\nu_e)$ total cross sections with scale
 uncertainties and corresponding $K$-factors at LHC 8TeV for various
 scenarios.} 
\label{tab:xsecWpH}
\end{table}

\begin{table}
\center
\begin{tabular}{lrlrlc}
 \hline
 scenario    & $\sigma_{\rm LO}$ & \hspace*{-3mm}(fb) 
             & $\sigma_{\rm NLO}$ & \hspace*{-3mm}(fb) & $K$ \\
 \hline
 $0^+$(SM)   & 22.46(1) & \hspace*{-3mm}${}^{+0.0\perc}_{-0.6\perc}$  
             & 29.86(3) & \hspace*{-3mm}${}^{+2.3\perc}_{-1.8\perc}$ 
             & 1.33 \\
 $0^+$(HD)   & 7.009(5) & \hspace*{-3mm}${}^{+1.4\perc}_{-1.7\perc}$
             & 9.355(9) & \hspace*{-3mm}${}^{+1.9\perc}_{-1.3\perc}$ 
             & 1.34 \\
 $0^+$(HDder)& 145.7(1) & \hspace*{-3mm}${}^{+4.1\perc}_{-3.9\perc}$
             & 193.8(1) & \hspace*{-3mm}${}^{+2.1\perc}_{-1.9\perc}$ 
             & 1.33 \\
 $0^+$(SM+HD)& 57.90(5) & \hspace*{-3mm}${}^{+2.8\perc}_{-2.9\perc}$
             & 77.31(8) & \hspace*{-3mm}${}^{+1.8\perc}_{-1.6\perc}$ 
             & 1.34 \\
 $0^-$(HD)   & 4.151(3) & \hspace*{-3mm}${}^{+2.5\perc}_{-2.6\perc}$
             & 5.550(5) & \hspace*{-3mm}${}^{+1.7\perc}_{-1.4\perc}$ 
             & 1.34 \\
 $0^\pm$(HD) & 5.583(4) & \hspace*{-3mm}${}^{+1.8\perc}_{-2.0\perc}$
             & 7.445(7) & \hspace*{-3mm}${}^{+1.8\perc}_{-1.3\perc}$ 
             & 1.33 \\
 \hline
\end{tabular}
\caption{Same as table~\ref{tab:xsecWpH}, but for 
 $pp\to H(W^-\to e^-\bar\nu_e)$.}
\label{tab:xsecWmH}
\end{table}

\begin{table}
\center
\begin{tabular}{lrlrlc}
 \hline
 scenario    & $\sigma_{\rm LO}$ & \hspace*{-3mm}(fb) 
             & $\sigma_{\rm NLO}$ & \hspace*{-3mm}(fb) & $K$ \\
 \hline
 $0^+$(SM)   & 10.13(1) & \hspace*{-3mm}${}^{+0.0\perc}_{-0.5\perc}$  
             & 13.24(1) & \hspace*{-3mm}${}^{+2.2\perc}_{-1.7\perc}$ 
             & 1.31 \\
 $0^+$(HD)   & 2.638(2) & \hspace*{-3mm}${}^{+1.4\perc}_{-1.7\perc}$  
             & 3.461(3) & \hspace*{-3mm}${}^{+1.9\perc}_{-1.3\perc}$ 
             & 1.31 \\
 $0^+$(HDder)& 48.61(4) & \hspace*{-3mm}${}^{+4.2\perc}_{-3.9\perc}$  
             & 63.59(5) & \hspace*{-3mm}${}^{+2.1\perc}_{-1.9\perc}$ 
             & 1.31 \\
 $0^+$(SM+HD)& 19.95(1) & \hspace*{-3mm}${}^{+3.1\perc}_{-3.1\perc}$  
             & 26.24(2) & \hspace*{-3mm}${}^{+1.8\perc}_{-1.6\perc}$ 
             & 1.32 \\
 $0^-$(HD)   & 1.480(1) & \hspace*{-3mm}${}^{+2.6\perc}_{-2.7\perc}$  
             & 1.952(1) & \hspace*{-3mm}${}^{+1.7\perc}_{-1.5\perc}$ 
             & 1.32 \\
 $0^\pm$(HD) & 2.061(1) & \hspace*{-3mm}${}^{+1.9\perc}_{-2.0\perc}$
             & 2.705(2) & \hspace*{-3mm}${}^{+1.8\perc}_{-1.3\perc}$ 
             & 1.31 \\
\hline
\end{tabular}
\caption{Same as table~\ref{tab:xsecWpH}, but for 
 $pp\to H(Z\to e^+e^-)$.} 
\label{tab:xsecZH}
\end{table}

\begin{figure*}
 \center 
 \includegraphics[width=0.48\textwidth,clip, trim = 90 210 50 160]{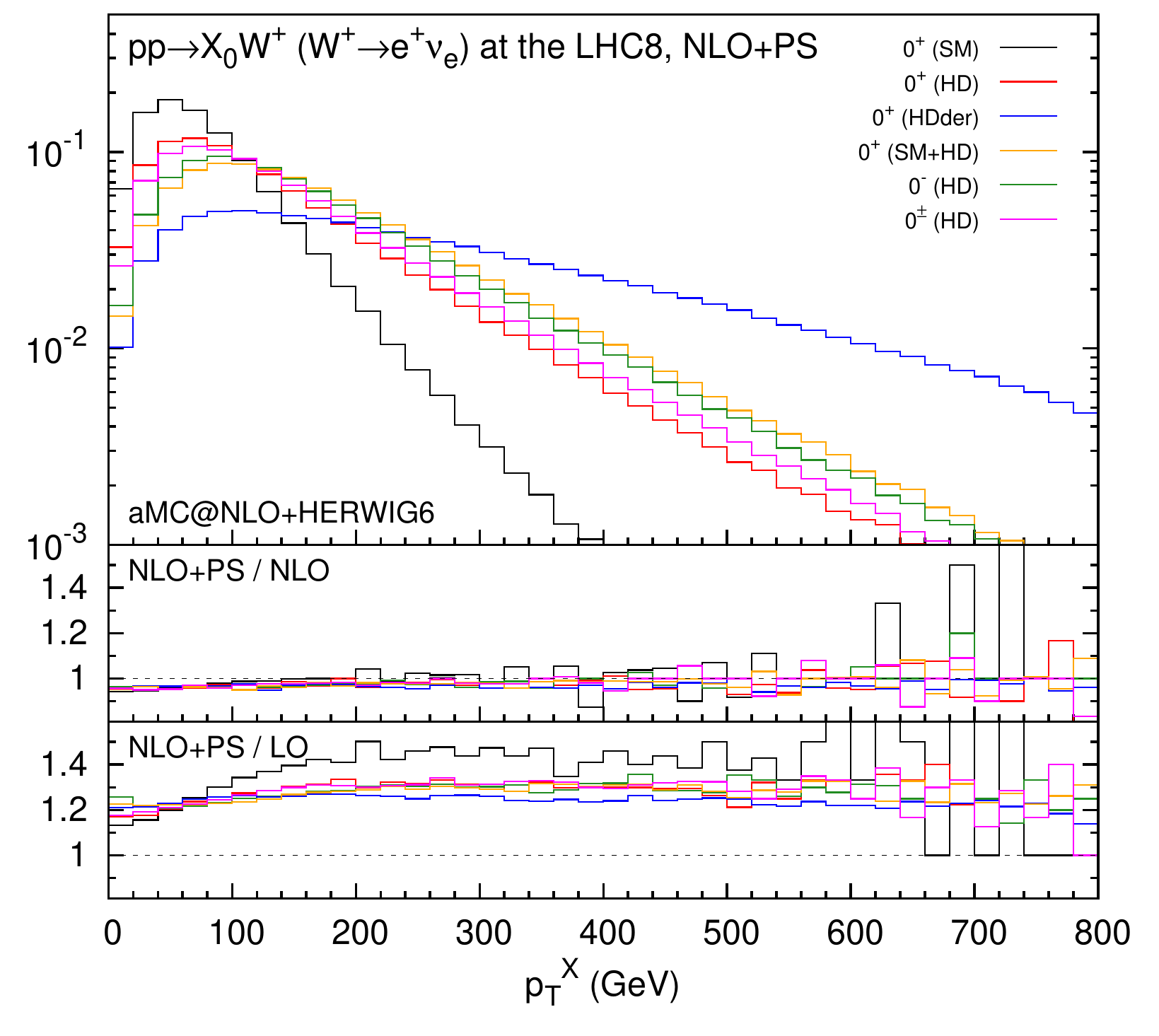}
 \includegraphics[width=0.48\textwidth,clip, trim = 90 210 50 160]{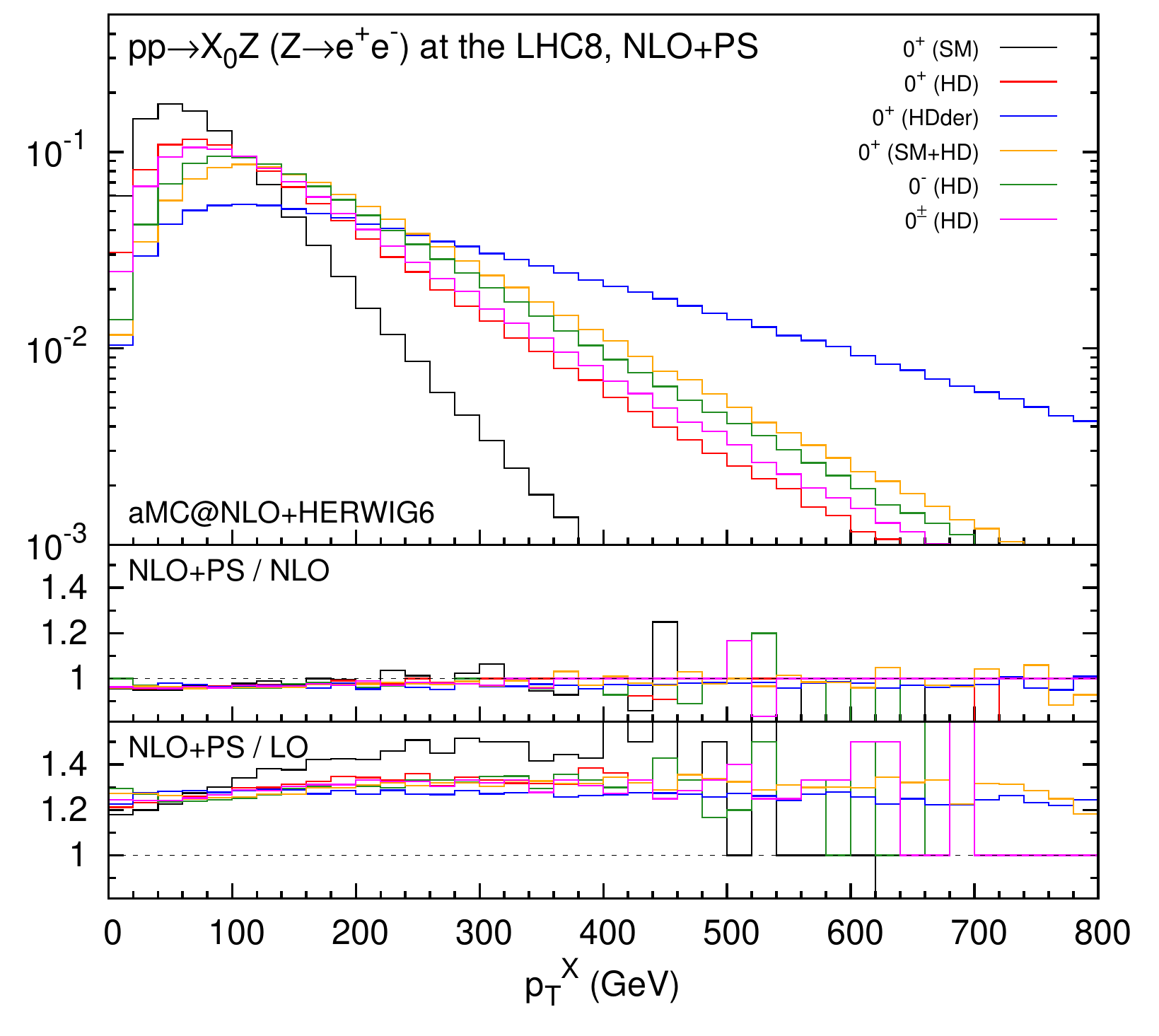}\\
 \includegraphics[width=0.48\textwidth,clip, trim = 90 210 50 160]{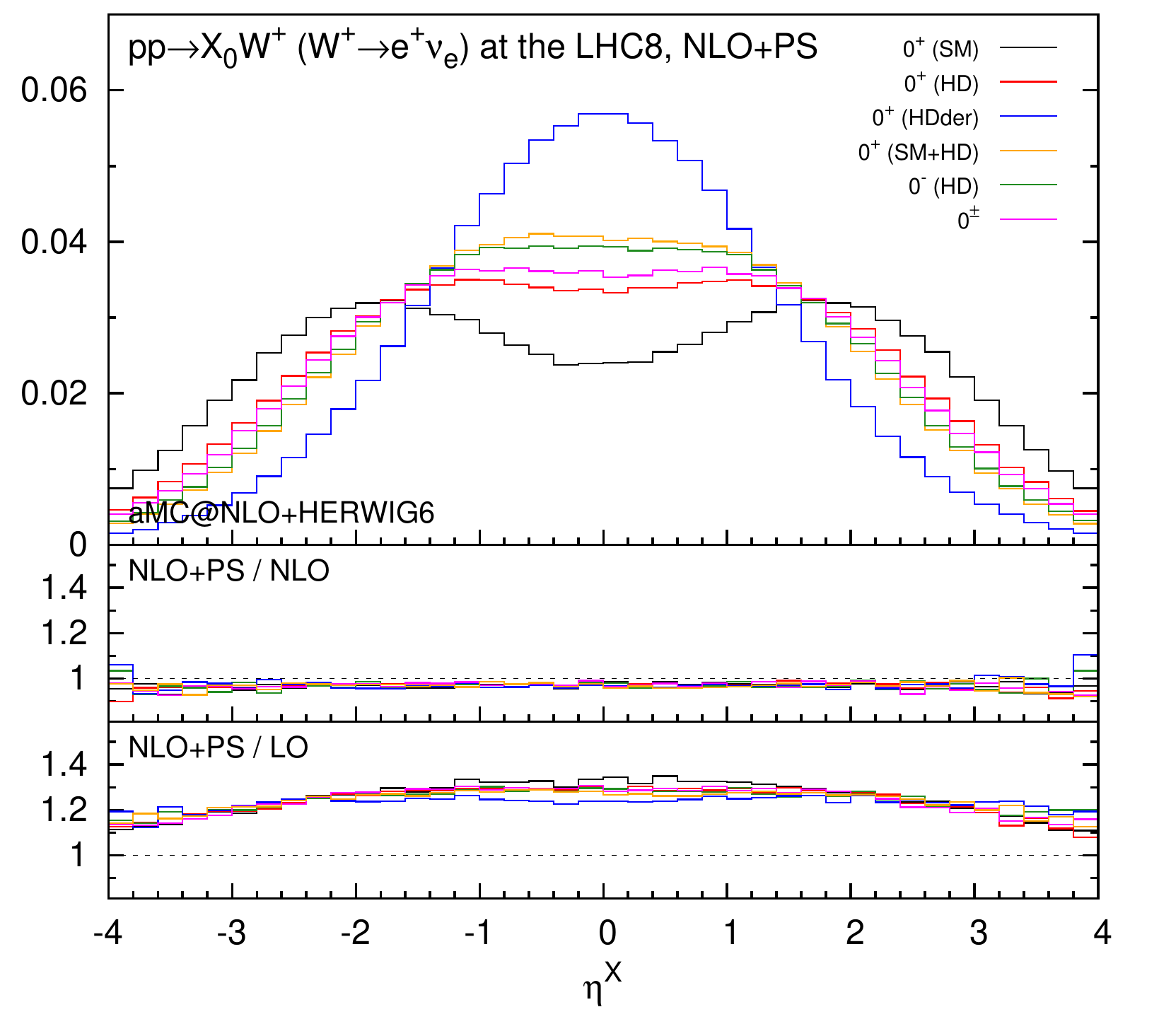}
 \includegraphics[width=0.48\textwidth,clip, trim = 90 210 50 160]{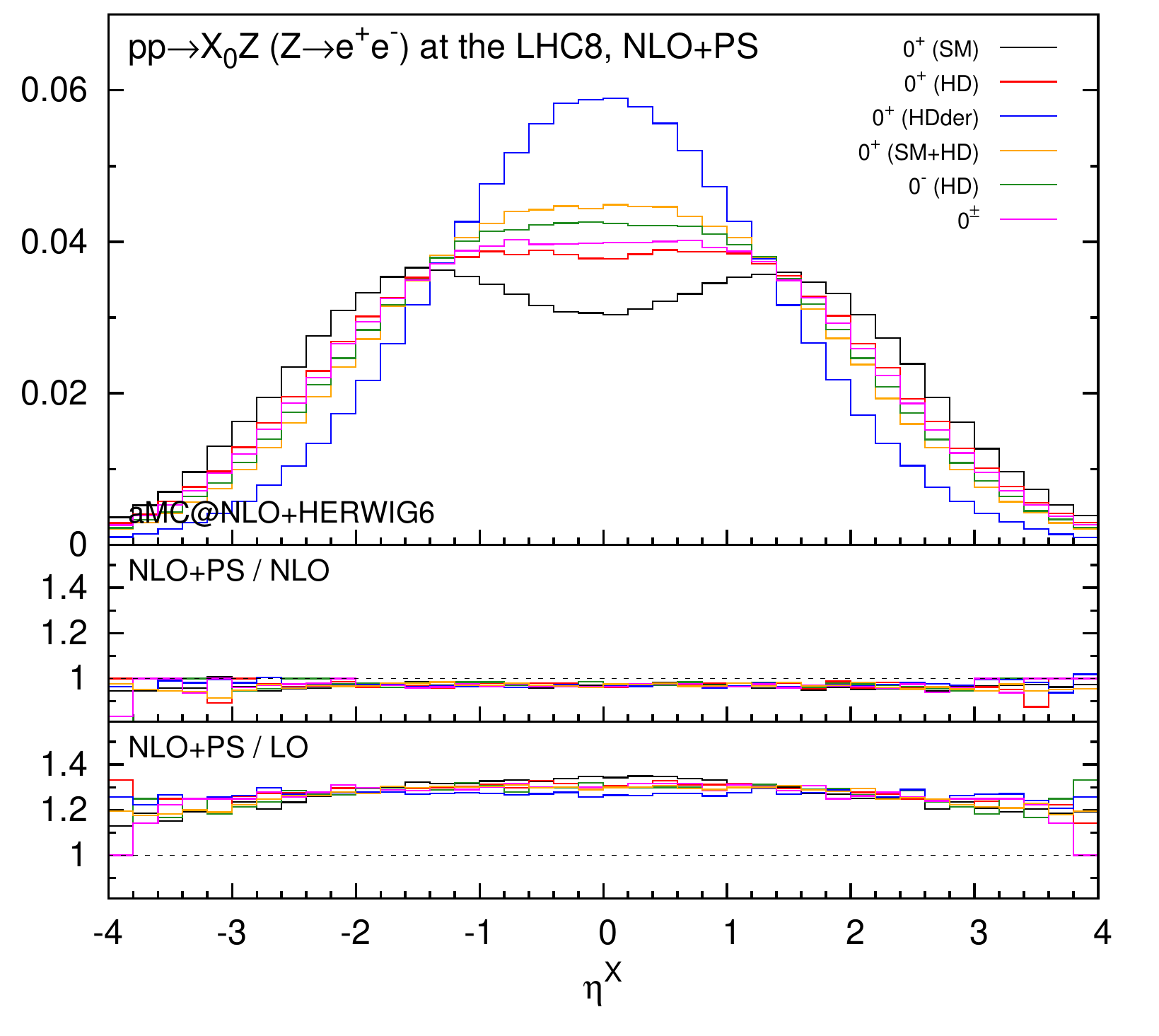}\\
 \includegraphics[width=0.48\textwidth,clip, trim = 90 210 50 160]{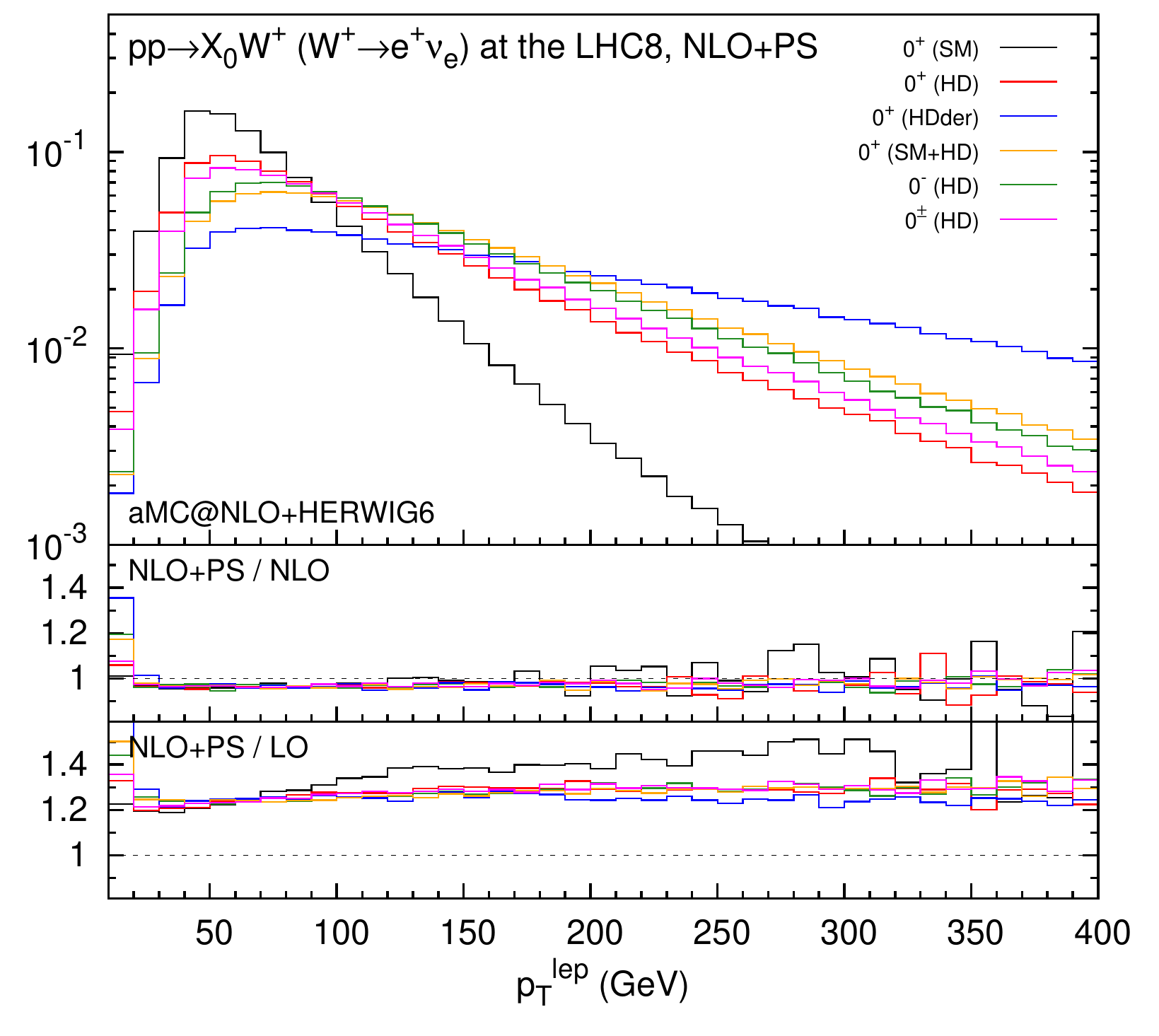}
 \includegraphics[width=0.48\textwidth,clip, trim = 90 210 50 160]{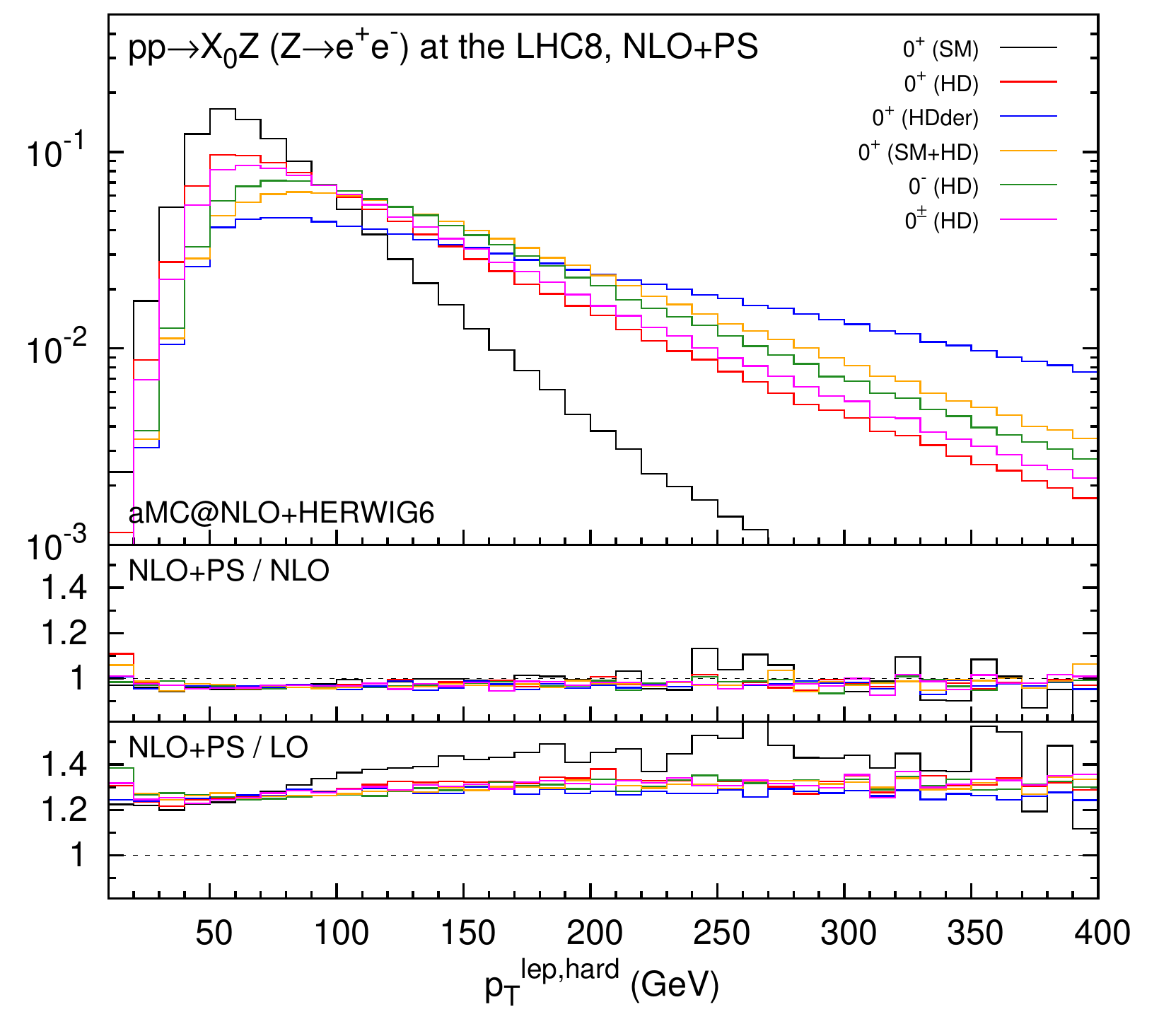}
 \caption{Distributions for $p_T^X$, $\eta^X$, and
 $p_T^{\ell}$ in $W^+H$ (left) and in $ZH$ (right) production with the
 acceptance cuts for the lepton(s). The histograms
 in the main plots are normalized to unity.}   
 \label{fig:VH}
\end{figure*}

We then show, fig.~\ref{fig:VH}, distributions for the several inclusive
variables with minimal cuts on the charged lepton(s): 
\begin{align}
 p_T^\ell>10~{\rm GeV}\,,\quad |\eta^\ell|<2.5\,,
\end{align}
for $W^+H$ and $ZH$ production (distributions for $W^-H$ are very
similar to $W^+H$ and we do not display them). 

\begin{figure*}
 \center 
 \includegraphics[width=0.48\textwidth,clip, trim = 90 210 50 160]{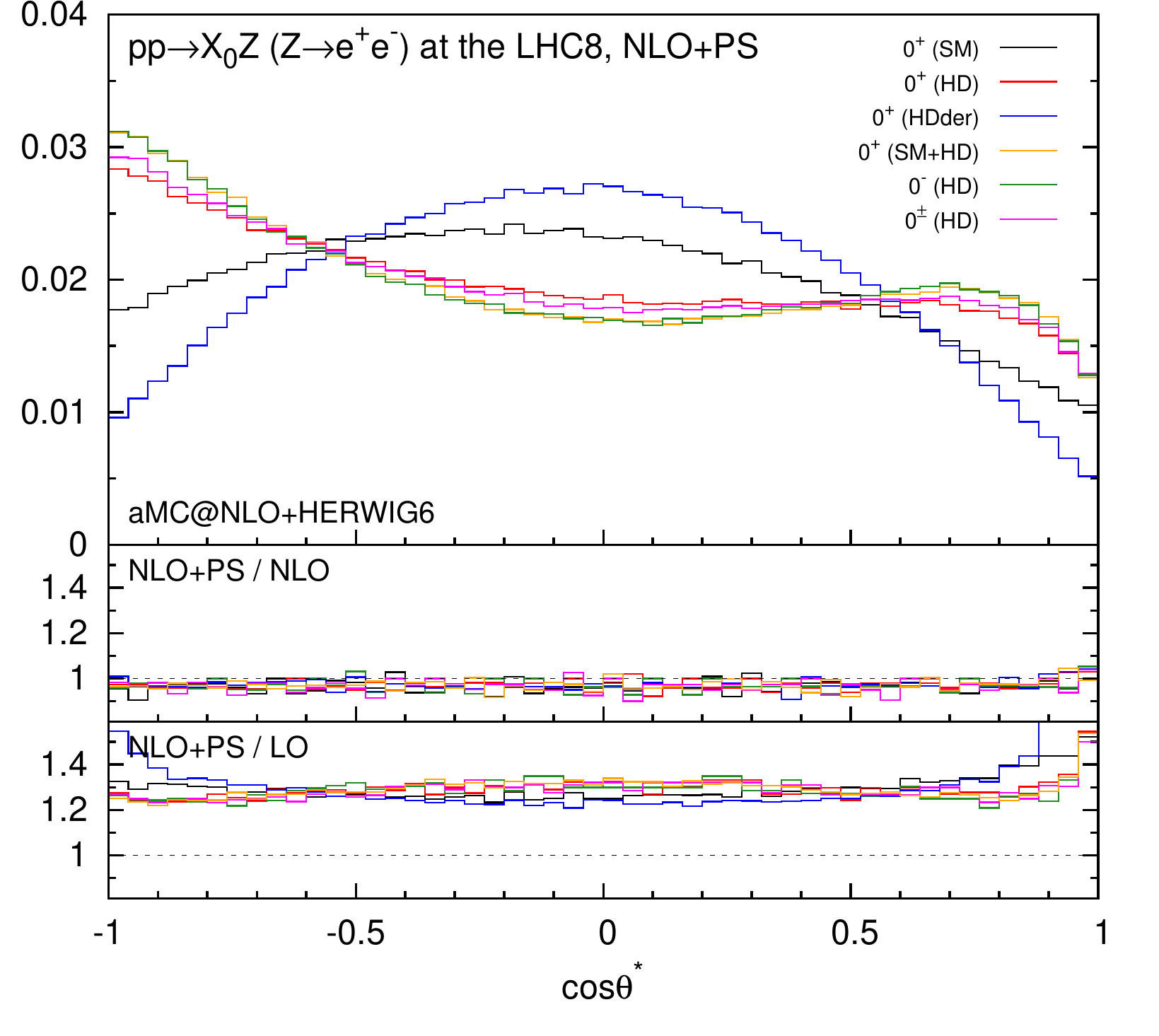}
 \includegraphics[width=0.48\textwidth,clip, trim = 90 210 50 160]{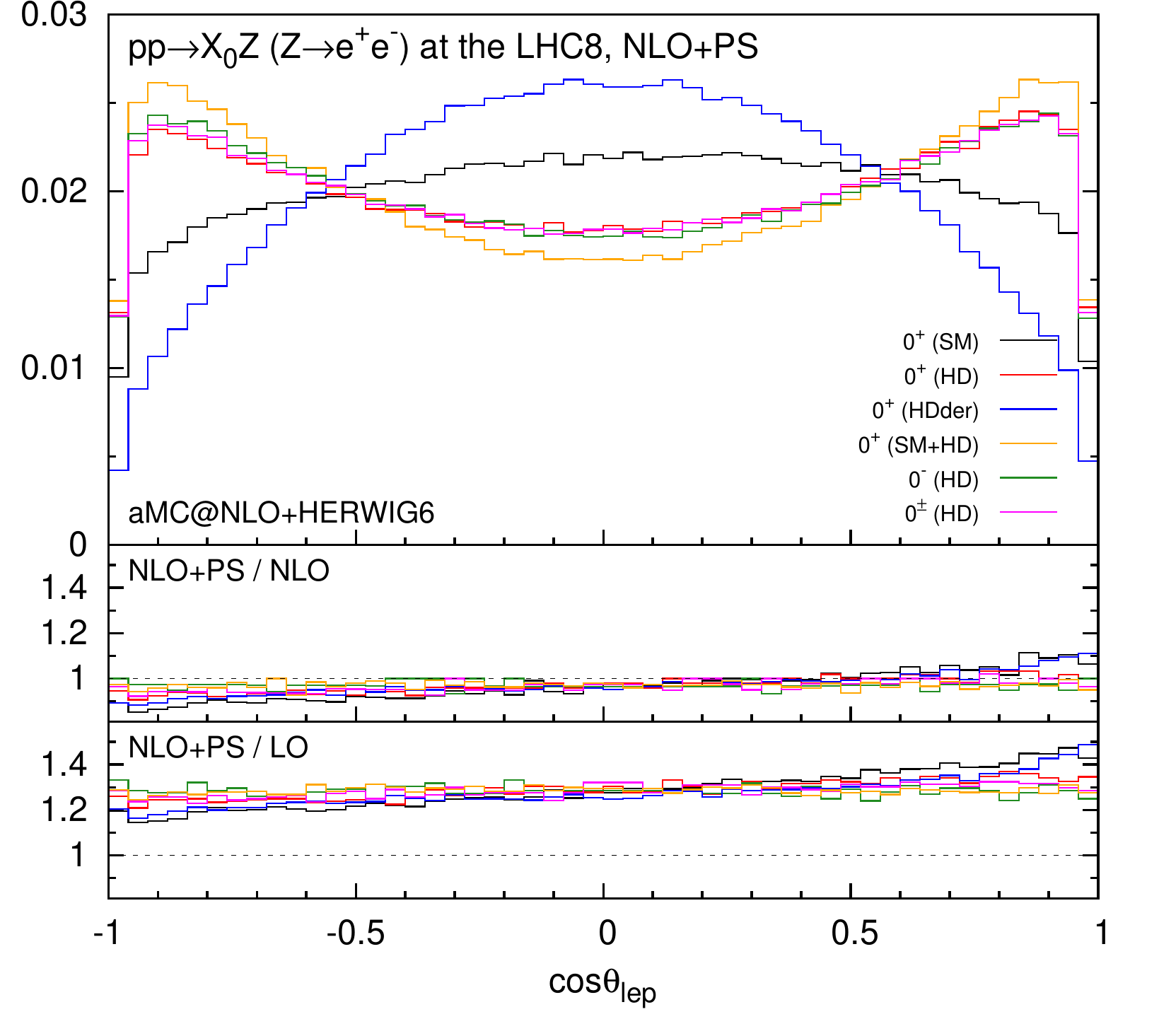}
 \caption{Distributions for $\cos\theta^*$ and
 $\cos\theta_{\ell}$ in $ZH$ with the acceptance cuts for the leptons.
 The histograms in the main plots are normalized to unity.} 
 \label{fig:ZH}
\end{figure*} 

The results for $W$ and $Z$ display very similar features. The scenarios
that include contributions from higher dimensional operators show harder
$p_T$ spectra. This is even more pronounced in the case of the derivative
operator (HDder). This fact is also reflected in the shape of rapidity
distributions, {\it i.e.}, the harder $p_T$ spectra correspond to more
central rapidity for the VH scattering.

As in sect.~\ref{sec:vbf}, the ratios of NLO+PS to LO (NLO) results are 
presented in the lowest (middle) inset in fig.~\ref{fig:VH}.
NLO+PS effects are quite important when compared with fixed-order
LO predictions, and, in many cases, they cannot be accounted for by
applying an overall $K$-factor. Conversely, NLO+PS distributions are in
almost perfect agreement with fixed-order NLO predictions, witnessing
small effects genuinely due to the parton shower. 

In fig.~\ref{fig:ZH} we show the polar angle distributions in $ZH$
production. $\cos\theta^*$ is defined as an angle between the
intermediate $Z^*$ momentum and the reconstructed $Z$ in the
$Z^*$ rest frame, while $\cos\theta_{\ell}$ is the lepton angle along
the $Z$ momentum in the $Z$ rest frame. In this case, NLO+PS corrections
do not affect the $\cos{\theta^*}$ distributions significantly, while
those of $\cos{\theta_\ell}$ are mildly modified. We note that the
asymmetry of the $\cos\theta^*$ distribution is due to the cuts on the
leptons.

\section{Summary}\label{sec:summary}

We have studied higher order QCD effects for various spin-0 hypotheses
in VBF and VH production, obtained in a fully automatic way via the
model implementation in {\sc FeynRules} and event generation at NLO
accuracy in {\sc MadGraph5\_aMC@NLO} framework. Our approach to Higgs
characterisation is based on an EFT that takes into account all relevant
operators up to dimension six written in terms of fields above the EWSB
scale and then expressed in terms of mass eigenstates ($W,Z,\gamma$, and
$H$). 

We have presented illustrative distributions obtained by interfacing NLO
parton-level events to {\sc HERWIG6} parton shower. NLO corrections
improve the predictions on total cross sections by reducing the scale
dependence. In addition, our simulations show that NLO+PS effects need
to be accounted for to make accurate predictions on the kinematical
distributions of the final state objects, such as the Higgs and the jet
distributions.  

We look forward to the forthcoming LHC experimental studies employing
the EFT approach and NLO accurate simulations to extract accurate
information on possible new physics effects in Higgs physics.

\section*{Acknowledgments}

We would like to thank all the members of Higgs Cross Section Working
Group for their encouragement in pursuing the Higgs Characterisation
project. 
We also thank Stefano Frixione for helpful comments on the draft.

This work has been performed in the framework of the ERC grant 291377
``LHCtheory: Theoretical predictions and analyses of LHC physics:
advancing the precision frontier'' 
and it is supported in part by the Belgian Federal Science Policy Office
through the Interuniversity Attraction Pole P7/37. 
The work of FM is supported by the IISN ``MadGraph'' convention
4.4511.10, the IISN ``Fundamental interactions'' convention 4.4517.08.
KM is supported in part by the Strategic Research Program ``High Energy
Physics'' and the Research Council of the Vrije Universiteit Brussel.  
The work of MZ is partially supported by the Research Executive Agency
(REA) of the European Union under the Grant Agreement number 
PITN-GA-2010-264564 (LHCPhenoNet).

\bibliography{library}
\bibliographystyle{JHEP}

\end{document}